\documentclass[12pt,preprint]{aastex}

\shorttitle{A Study of the Near-Ultraviolet Spectrum of Vega}
\shortauthors{A. Garc\'ia-Gil et al.}

\begin{document}

\title{A Study of the Near-Ultraviolet Spectrum of Vega}

\author{Alejandro Garc\'{\i}a-Gil}
\affil{Instituto de Astrof\'{\i}sica de Canarias, c/V\'{\i}a L\'actea s/n,
E-38200, La Laguna, Tenerife, Spain}
\email{agg@ll.iac.es}

\author{Ram\'on J. Garc\'{\i}a L\'opez}
\affil{Instituto de Astrof\'{\i}sica de Canarias, c/V\'{\i}a L\'actea s/n,
E-38200, La Laguna, Tenerife, Spain}
\affil{Departamento de Astrof\'{\i}sica, Universidad de La Laguna, Avenida
Astrof\'{\i}sico Francisco S\'anchez s/n, E-38206, La Laguna, Tenerife,
Spain}
\email{rgl@ll.iac.es}

\author{Carlos Allende Prieto}
\affil{McDonald Observatory and Department of Astronomy, The University of
Texas, Austin, TX 78712-1083, USA}
\email{callende@hebe.as.utexas.edu}

\and

\author{Ivan Hubeny}
\affil{AURA/NOAO, Tucson, AZ 85726-6731, USA}
\email{hubeny@tlusty.gsfc.nasa.gov}

\shorttitle{A Study of the Near-Ultraviolet Spectrum of Vega}
\shortauthors{A. Garc\'ia-Gil et al.}

\begin{abstract}
UV, optical, and near-IR spectra of Vega have been combined to test our
understanding of stellar atmospheric opacities, and to examine the
possibility of constraining chemical abundances from low-resolution UV fluxes.
We have carried out a detailed analysis assuming Local Thermodynamic 
Equilibrium (LTE) to identify the most important
contributors to the UV continuous opacity: H, H$^{-}$, C I, and Si II. 
Our analysis also assumes that Vega is spherically symmetric and its atmosphere 
is well described with the plane parallel approximation. Comparing 
observations and computed fluxes we have been able to 
discriminate between two different flux scales that have been 
proposed, the IUE-INES and the HST scales, favoring the latter. The effective
temperature and angular diameter derived from the analysis of 
observed optical and near-UV spectra are in very good agreement with
previous determinations based on different techniques.
The silicon abundance is poorly constrained by the UV observations of 
the continuum and strong lines, but
the situation is more favorable for carbon and the abundances inferred from the 
UV continuum and optical absorption lines are in good agreement. Some spectral 
intervals in the UV spectrum of Vega that the calculations do not reproduce 
well are likely affected by deviations from LTE, but we conclude that our 
understanding of UV atmospheric opacities is fairly complete for early A-type stars.  
\end{abstract}

\keywords{stars: atmospheres --- stars: fundamental parameters ---
stars: abundances --- stars: individual (\objectname{Vega}) --- 
techniques: spectrophotometry --- ultraviolet: general}

\section{Introduction} 

Understanding the formation of the UV spectrum of individual 
stars is essential in order  
to quantitatively determine  stellar abundances for  
a number of chemical elements that
 only produce measurable features in the UV domain such as B, Be, or
the neutron capture elements Pb and Ge. A proper understanding
of the spectra of individual stars is also crucial 
 to interpret  integrated light from galaxies and other stellar systems.
The UV radiation field,  
 besides affecting directly the abundances obtained from UV lines,
has an important effect on
 the construction of atmospheric models through the
energy balance. Theoretical atmospheric structures 
computed under the assumptions of radiative equilibrium and local
thermodynamical equilibrium are routinely employed in the analysis of
spectra from astronomical objects. Consequently, an error in the UV opacities
translates into wrongly derived physical properties, namely, chemical
composition and atmospheric parameters 
for individual stars, or ages for galaxies.

A possible UV opacity source missing from the calculations 
 has been proposed to explain an early reported 
failure to match the observed solar UV spectrum. 
It has been suggested that either 
line \citep[e.g.][ Vernazza, Avrett \& Loeser 1976]{Ho70}
or continuum \citep[][ Bell, Balachandran \& Bautista 2001]{BPT94} 
metal absorption could provide that extra opacity. However, 
several recent studies of UV spectra of other stars suggest that there 
is a reasonable agreement between 
theoretical models \citep{K92} and observations
\citep[e.g.][ Fitzpatrick \& Massa 1999, Peterson et al. 2001] {APL00}.
The analysis of the solar spectrum by \citet{Be01} concluded  
that even when using the most recent calculations of line and continuum
opacities, classical model atmospheres predict too much UV flux. 
Nonetheless, this result has been challenged by independent calculations
by \citet{AP03}, who found that
LTE modeling can reproduce closely the observed solar 
flux at wavelengths
longer than about 2600 \AA, and uncertainties in the atomic line data 
 fully account for the differences between computed and observed fluxes.
High quality observations of stars with different atmospheric parameters
than the Sun present a promising way to solve the controversy.

There is  a second 
controversy that affects the absolute flux scale of UV observations.
 The  most recent calibration of the  
 {\it International Ultraviolet
Explorer} (IUE) archive has been dubbed INES ({\it IUE Newly
Extracted Spectra}), and will be discussed more deeply in the following
section. IUE obtained the largest available 
collection of astronomical UV spectra.
There are at least two more calibrations of IUE, both taking spectra to the
HST flux scale, which
is 7.2 \% lower than the INES scale
 \citep{GRCW01}. These calibrations will be
explained in Sect. 2. Clarifying which flux scale should be adopted 
is a must to properly interpret observations.

In this paper
we deal with Vega ($\alpha$ Lyr, HD 172167, A0V). This star is not straightforward to
model, because it has a dust and gas disk \citep{W83,2002ApJ} which produces an IR
flux excess; it is a fast rotator seen pole-on \citep{G94}; and it is
possibly variable \citep{V89}. Moreover, it has a peculiar abundance pattern,
 similar to that of $\lambda$ Boo stars \citep{I98}, that is, with a
relative 
underabundance of Fe-like elements with respect to the Sun, but solar
proportions of C, N, O, and S. 
Despite these difficulties, Vega is one of the few dwarf stars that can be
used to test detailed calculations of UV irradiances:
both precise UV spectrophotometry 
 and independent measurements of its angular diameter are available. 
Visible high resolution and 
high signal-to-noise ratio (S/N) spectra of Vega are also readily obtained
from the ground. Besides,  the
disk is expected to make no detectable 
contribution to the star's visible and UV spectrum, and it has been found that 
the influence of fast
rotation is only important for the shape of spectral lines, with a 
small effect on the continuum limited to the spectral
region in the vicinity of the Balmer jump \citep{G94}. 

Our analysis puts to the test calculations of UV irradiances based 
on classical (Kurucz 1992) model atmospheres and modern line and continuous 
opacities. In addition to UV spectrophotometry of the star, 
we also use optical spectrophotometry and 
high-dispersion optical spectra to constrain the
chemical abundances and narrow down the ranges of the involved 
model parameters.
In \S~2, we present a description of the observations employed. In \S~3,
we describe 
 the analysis carried out and the results
obtained, whose consequences are discussed in \S~4.

\section{Description of the observations}

 We have considered three versions 
 of the average low-dispersion IUE spectrum of Vega:
 
 \begin{itemize}
 
  \item INES; this is 
  not a calibration of IUE data processed with NEWSIPS ({\it New Spectroscopic
  Image Processing System}; see Garhart et al. 1997), but a new 
full processing of
  the raw data, improving the spectrum extraction to diminish the error sources 
  present in NEWSIPS (improving the noise model, the extraction method, the
  homogenization of the wavelength scale, and the flagging of the absolute
  calibrated extracted spectra). The relative fluxes
  were calibrated using a pure H (DA) white dwarf,  G191B2B, whose
  virtually 
featureless spectrum is relatively simple to model. 
The absolute flux scale relies on that of the UV satellite OAO-2 
\citep{GRCW01}.
  
  \item Bohlin's calibration \citep{Bo96}, obtained from the CALSPEC Web page 
 \footnote{\tt http://kahuna.stsci.edu/instruments/observatory/cdbs/calspec.html}; 
 8 standard stars (4 DA's among them) are used to calibrate FOS ({\it Faint Object 
 Spectrograph}) observations, formerly onboard HST. By comparing 
 their FOS and IUE fluxes, a correction factor is determined, which is applied 
 to each wavelength of a IUESIPS (the earlier version of NEWSIPS) spectrum of 
 Vega to place the data in the HST flux scale.
  This set of standard spectra has been updated by new STIS and NICMOS 
  observations \citep[both onboard  HST; ][] {BDC01}\footnote{The spectrum
  we use in this article is not the one obtained with STIS and recently published 
  \citep{Bo04}, but the earlier one, since this one was not yet published at the 
  time this study was carried out}.  The zero
 level of the flux scale is tied to the average over the V band for Vega as
  measured by \citet{Ha85}.
  
  \item Massa \& Fitzpatrick's calibration \citep[][ hereafter M\&F] {MF00}; it
  takes into account the systematic errors in NEWSIPS spectra
with the observation
  time and the temperature of the amplifying camera (THDA), and applies a
  correction factor to place the data into the HST scale.
  
 \end{itemize}
 
 The three versions above are based on low-dispersion large-aperture
 IUE spectra, but
 the particular individual spectra considered in the average, 
 and some parts of the
 processing (chiefly, the flux calibration) differ among them. 
  These spectra have a large spectral coverage with a 
  resolution of $\sim$ 6 \AA. 
We combined several IUE 
spectra from the INES and MAST\footnote{Multi-mission
Archive at Space Telescope} servers to produce the
 INES and M\&F versions, respectively. The list of individual spectra 
 is provided in Table \ref{IUE_cam}. Before combining the MAST-IUE spectra to produce
 the  M\&F version, the fluxes were processed with the software kindly
 provided by these authors. Bohlin's spectrum was directly obtained from
 the STScI web pages, as provided by R. Bohlin\footnote{URL is  
{\tt ftp://ftp.stsci.edu/cdbs/cdbs2/calspec/}}.

\begin{figure}
   \epsscale{.80}
 \plotone{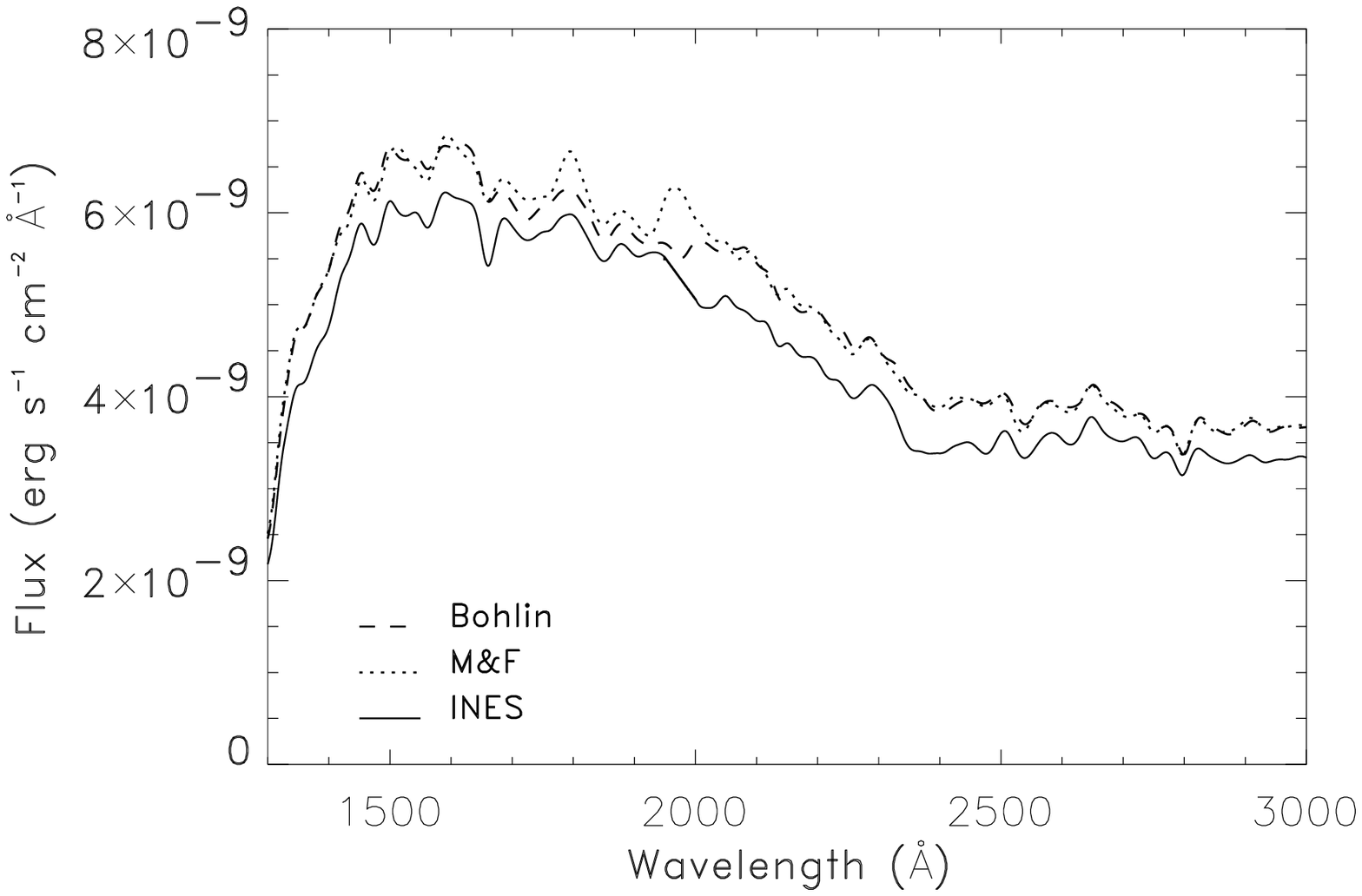}
\caption{Comparison of the observed fluxes in the UV obtained with 
different calibrations of the IUE satellite smoothed by convolution with a 
Gaussian profile 
with FWHM = 25 \AA\ (see text for details).} 
 \label{fig:IUEcal}
\end{figure}

As we can see in Fig. \ref{fig:IUEcal}, Bohlin's
 and M\&F fluxes (which share the same flux scale) are $\sim$10\% higher 
 than the INES fluxes for Vega. This is higher, but roughly consistent within
 the error bars with the global
 differences reported by \citet{GRCW01} between FOS and IUE spectra for the INES reference 
 star G191B2B. However, they noticed an almost linear reduction
 from a difference of $\sim$7 \% at
 1500 \AA\  to $\sim$6 \% at 3000 \AA, while  we find
  larger differences for Vega 
 in the ranges 1500 to 1700 \AA\ and  2000 to 2500 \AA.
 Fig. 1 reveals that the M\&F flux is higher than 
  Bohlin's at
 $\sim$1800 and $\sim$2000 \AA. The feature in the M\&F version at
 $\sim$2000 \AA\ coincides with the 
 border between the short and long-wavelength cameras, where the
 uncertainty in the flux is higher.

We have also compared our calculations with high-resolution IUE spectra. These 
spectra have a large spectral coverage with a resolution $\sim 0.2$ \AA. Table 
\ref{IUE_cam} lists  the  high-resolution spectra considered, which were corrected 
in absolute flux to put them into the HST scale and then averaged to obtain the final 
one.

 We also compared Bohlin's and INES fluxes for
 the eight primary standard stars on which Bohlin's fluxes are based
 (G191B2B, GD153, GD71, HZ43, HZ44, BD +28 4211, BD +33 2642, and BD +75 325). 
 We used all the IUE spectra available for these stars from
the  INES database, selecting
 only those pixels labeled with good quality and with flux errors 
 lower than 20\%. Bohlin's flux is again higher than the 
 INES one at all wavelengths and for all the stars, with the exception
 of  the SWP spectrum   of BD +75 325. 
The mean relative difference in flux for the other 
seven stars is  $(F_{Bohlin}-F_{INES})/F_{INES} = 0.066 \pm 0.013$, 
 which is again consistent with the global differences
  reported by \citet{GRCW01}. 
When studying the  wavelength 
dependence, an increment of the difference is seen at $\sim$2000 \AA\ 
and from 2200 to 2500 \AA, where the difference is over 8\%, while 
in the other regions it is 5-7\%, quite consistent with the larger 
differences  observed in Vega's case between 2000 and 2500 \AA.

Visible spectrophotometry provides additional information that cannot
 be neglected.
We consider the spectrum compiled by \citet{Ha85} by combining six 
different sources of ground-based measurements.
This spectrum has a flux uncertainty of just about 1-2\% in the optical 
 and covers a wavelength range from 3300 to 10500 \AA. The 
 flux is tabulated in steps of 25 \AA. This is both 
 the bandwidth and step sizes
 reported by Terez (1982) and Arkharov \& Terez (1982), who measured 
 Vega's flux in the entire region critically evaluated by Hayes.
The other sources considered by Hayes provide fluxes measured over
 smaller spectral ranges with bandpasses from 10 to 100 \AA, 
 and approximate corrections for the bandwidth. 
Based on this information, 
 it seems therefore appropriate to compare these data to 
 calculated fluxes smoothed to a resolution of at least 50 \AA.

 In addition to the basic atmospheric parameters: effective temperature,
 surface gravity, and overall metallicity, 
the abundances of particular elements, whose ratio to iron 
may differ from solar, may affect the spectral energy distribution in the
 UV. Initial estimates of those abundances are necessary to perform
 a meaningful comparison of computed and observed UV fluxes.
 The analysis of optical spectral lines provides abundances for the
 most relevant elements. 
  
  We have used two different spectra, with different resolving power, 
  obtained by our group at McDonald Observatory (Mount Locke, 
 Texas), acquired with the {\it 2dcoud\'e} cross-dispersed echelle spectrograph 
 \citep{T95} coupled to the Harlan J. Smith 2.7m Telescope. The first spectrum 
 is part of the S$^4$N survey of nearby stars (Allende Prieto et al. 2004) and 
 is the average of observations secured in May 1 and 2, 2001.
 It has a resolving power (R $\equiv \frac{\lambda}{\delta
 \lambda}$) of $\sim 5 \times 10^4$, covering in full the optical range,
 extending into the near IR. As the spectrograph provides 
 full coverage only from about 3600 \AA\
 to $\sim$5100 \AA\ in a single exposure for this resolution, with
  increasingly larger gaps between redder orders, two overlapping settings were 
  used to circumvent the problem.  
The second spectrum was obtained at the focal station F1, and 
has R $\sim$ 173000, covers the range from $\sim$4400 to $\sim$7600 \AA\ with 
some small gaps, and was obtained in June 9--12, 2000. The spectral coverage is 
much smaller at F1, and 10 tilts of the grating/prisms were needed to achieve 
the final coverage. The CCD registered pieces of $\sim$16 adjacent orders.

 For both the high- and intermediate-resolution 
spectra, the detector was TK3, 
a thinned $2048\times2048$ Tektronix CCD with 24 $\mu$m pixels, and we used 
 grating E2, a 53.67 gr mm$^{-1}$ R2 echelle from Milton Roy Co. For the
 spectrum with R $\sim 6 \times 10^4$, we used
 slit \#4, which has a central width of 511 $\mu$m (or 
 approx. 1.2 arcsec on the sky). For the spectrum with R $\sim$ 173000, we used
 slit \#2, which has a central width of 145 $\mu$m, and partially masked the
 collimator. A careful
 data reduction was applied to both spectra, using  
 IRAF\footnote{IRAF is distributed by the National Optical Astronomy
 Observatories, which are operated by the Association of Universities for
 Research in Astronomy, Inc., under cooperative agreement with the National
 Science Foundation}, which consisted of overscan (bias) and
 scattered-light subtraction; flatfielding; extraction of one-dimensional
 spectra; wavelength calibration; and continuum normalization.

\section{Analysis}

 \subsection{Chemical species affecting the visible and the UV continuum}
 \label{absorp}
 
 A few particular elements may contribute significant opacity in the UV. Iron
 is known to dominate line absorption virtually for all 
 spectral types. Continuous absorption from other metals may also
 be significant or even dominant. The main contributors to the
UV continuous opacity change 
  with the spectral type and, therefore, it is important
 to evaluate which metals are relevant and provide reasonable estimates
 of their abundances to calculate realistic UV fluxes.
 The ability of an ion to contribute bound-free opacity depends on
 its atomic structure and its abundance.
 As a preliminary step, 
 ion abundances for the 30 first elements in the first three
 states of ionization were computed for a star like Vega, using the
 model parameters preferred by Castelli \& Kurucz (1994;  
 an effective temperature T$_{\rm eff}$ = 9550 K and gravity $\log g$ = 3.95)
 using Saha and Boltzmann equations. At this point we simply
 assumed solar photospheric 
 abundances \citep{AG89}. 
 
\begin{figure}
   \epsscale{.80}
\plotone{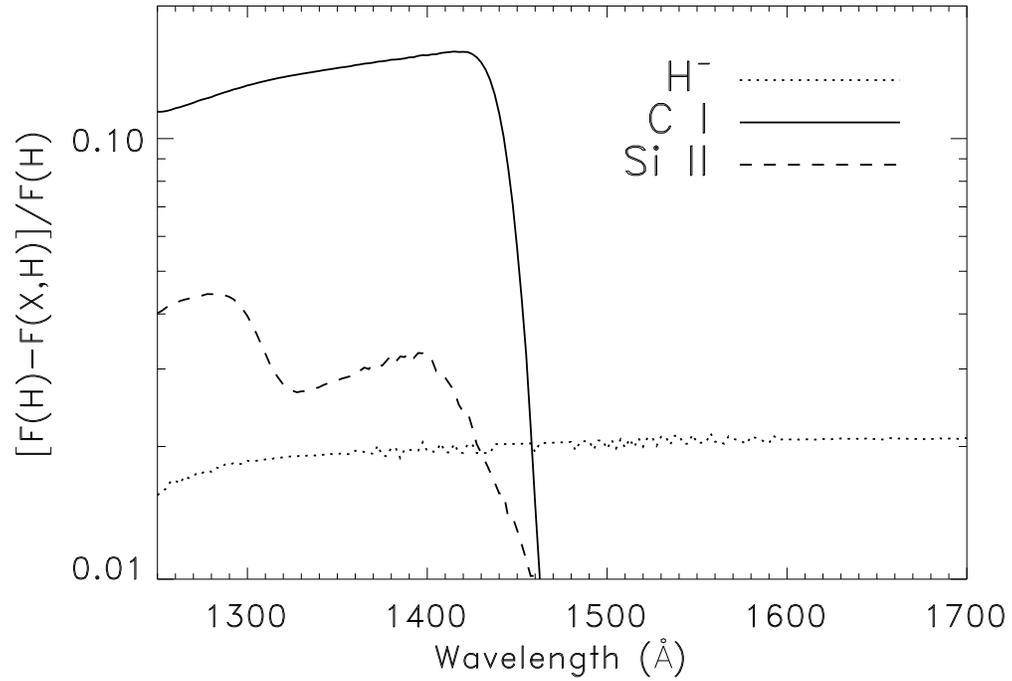} 
\caption{Relative difference between the flux computed taking into account
neutral H and each ion X explicitly (F(X,H)) and the flux obtained by just 
taking into account neutral H explicitly (F(H)). The change in neutral H
opacity is also considered in the line labeled as H$^-$, because the 
number of protons is constant, and therefore the abundance of neutral H 
diminishes when including H$^-$. Note the
importance of C\,{\scriptsize I} below 1450 \AA, and that of H$^-$ above this
wavelength.}
 \label{fig:cont_opac}
\end{figure}

 Next, we computed synthetic visible and near-UV spectra including continuous 
 opacity produced by only H (including bound-bound H opacity, i.e. hydrogen lines, 
 as part of continuous opacity) and each of the most abundant ions in Vega. 
 The calculations were carried out with the program SYNPLOT, a wrapper of SYNSPEC 
 \citep{HL00} for IDL. Bound-free opacities were obtained from TOPBASE \citep{Cu93} 
 and smoothed to dilute resonances whose exact frequencies are not accurately known
 (Allende Prieto et al. 2003). We studied all the ions with abundances relative to 
 H larger than 10$^{-6}$: He\,{\scriptsize I},
 C\,{\scriptsize I}, C\,{\scriptsize II}, N\,{\scriptsize I}, N\,{\scriptsize
 II}, O\,{\scriptsize I}, O\,{\scriptsize II}, Na\,{\scriptsize II},
 Mg\,{\scriptsize II}, Al\,{\scriptsize II}, Si\,{\scriptsize II},
 S\,{\scriptsize II}, Ca\,{\scriptsize II}, Fe\,{\scriptsize II} and
 Ni\,{\scriptsize II}.
 A contribution could be missed if it corresponds to an 
 ion with a low abundance but an unusually   
 high photoionization cross section, but we believe our list is complete.
 For these calculations, the atmospheric parameters
 of \citet{CK94} were used as well as solar abundances for each element, 
 except for Fe, for 
 which it was adopted [Fe/H] $= -0.38$, to be consistent with the model
 atmosphere: [M/H] = $-$0.5 \citep[see the discussion in][] {FM99}. 
 By comparing the spectra derived including continuous opacity due to 
 H plus any given ion to that obtained including only 
 neutral H continuous opacity,
 we deduce the relevance of each ion's continuous opacity to the observed
 fluxes. This is
illustrated in Fig. \ref{fig:cont_opac} for the most important contributors. 
 Hydrogen bound-free
 and line absorption dominate the opacity blueward of L$_{\alpha}$,
 shaping the near-UV and optical spectrum
 of Vega. 
 At wavelengths longer than about 1500 \AA, neutral hydrogen and the H$^{-}$
 ion dominate the total bound-free opacity.
 Si and C contribute significantly to the continuous opacity 
 between 1200 and 1450 \AA.
 He\,{\scriptsize I} would also appear
  in the figure below 1400 \AA\ with a maximum of $\sim$1.5 \%, but it has not
  been included for clarity.
 
\begin{figure}
   \epsscale{.80}
\plotone{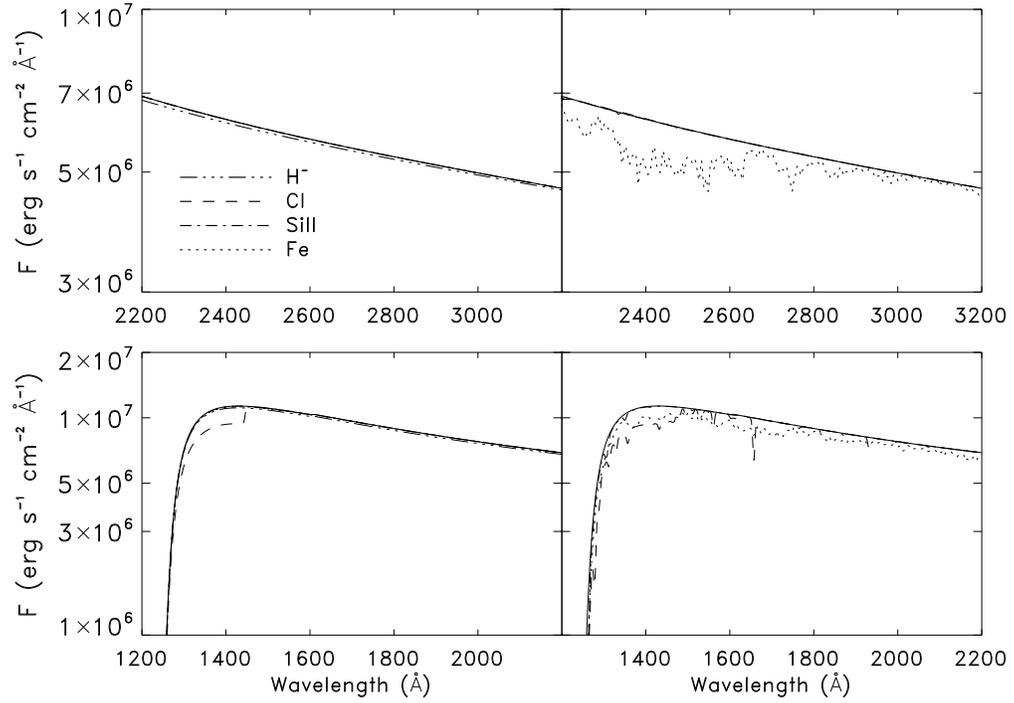} 
\caption{{\it Left}: Emergent flux when only the continuum absorption
produced by H$^-$, C\,{\scriptsize I}, Si\,{\scriptsize II} and Fe is
considered. {\it Right}: Emergent flux when the total (continuum+metal
line) opacity is considered. The curve corresponding to H$^-$ has been omitted
in the right panels for clarity. The solid line is the flux when no opacity is
considered.}
 \label{fig:opac}
\end{figure}

Fig. \ref{fig:opac} shows the
  continuum and line absorption due to each ion. 
Fe\,{\scriptsize I} lines swamp the UV spectrum of solar-like stars. At the 
temperatures in Vega's photosphere, iron is almost completely ionized
(N$_{\rm Fe I}/$N$_{\rm FeII} \sim 10^{-4}$), and the atomic line
absorption is mainly produced by Fe\,{\scriptsize II} (around half the total 
absorption of lines and most of the absorption due to Fe lines). 
Molecules are virtually inexistent at these warm temperatures.
We considered different sources for the atomic 
line data, finding that a linelist
based on data 
compiled by Kurucz and distributed with SYNPLOT provided the
best results. The agreement with the observations was only
marginally improved when
updating Stark  
line damping parameters extracted from VALD\footnote{Van der Waals parameters are similar,
so we have maintained them} \citep{K99}. All the fluxes shown in Fig. \ref{fig:opac} include a contribution to the total opacity 
by incoherent electron scattering. 
Although Thomson scattering is wavelength
independent, the vertical structure of the atmosphere makes its 
role maximum at about 1300 \AA, where it 
effectively reduces the emerging flux by about 4 \%, and
over 1 \% for $\lambda < 2000$ \AA.

In conclusion, to model the UV spectrum of a star like Vega it is necessary to 
account for H, C, Si, and Fe opacities in as much detail as possible.

 \subsection{Estimation of T$_{\rm eff}$ and $\theta$ by using visible and near-UV
 regions}
 \label{pars}
 
\citet{APL99} estimated a set of fundamental parameters for a large sample
of nearby stars, using Hipparcos parallaxes to accurately constrain their positions
in the color-magnitude diagram. Comparison with evolutionary calculations by
\citet{Be94} allowed them to provide masses, radii, gravities and effective
temperatures for 17,219 stars. Their estimate for Vega provides: $\log g = 3.98 \pm
0.02$. Given the limited effect of gravity on
 the UV spectrum, we will adopt this value and attempt to determine
 the effective temperature and angular diameter from the observations
 in the optical and near-infrared. This can be achieved by comparing the
 computed emitted flux from Kurucz model grids (which have $\log g$ and T$_{\rm eff}$
 as the independent parameters), transformed to computed received flux by using
 the angular diameter, to the observed flux. In these regions, metal absorption
 is modest, which allows us to study these two parameters fairly
 independent of the details of the chemical composition.

\begin{figure}
\begin{center}
{\par \includegraphics[scale=0.7]{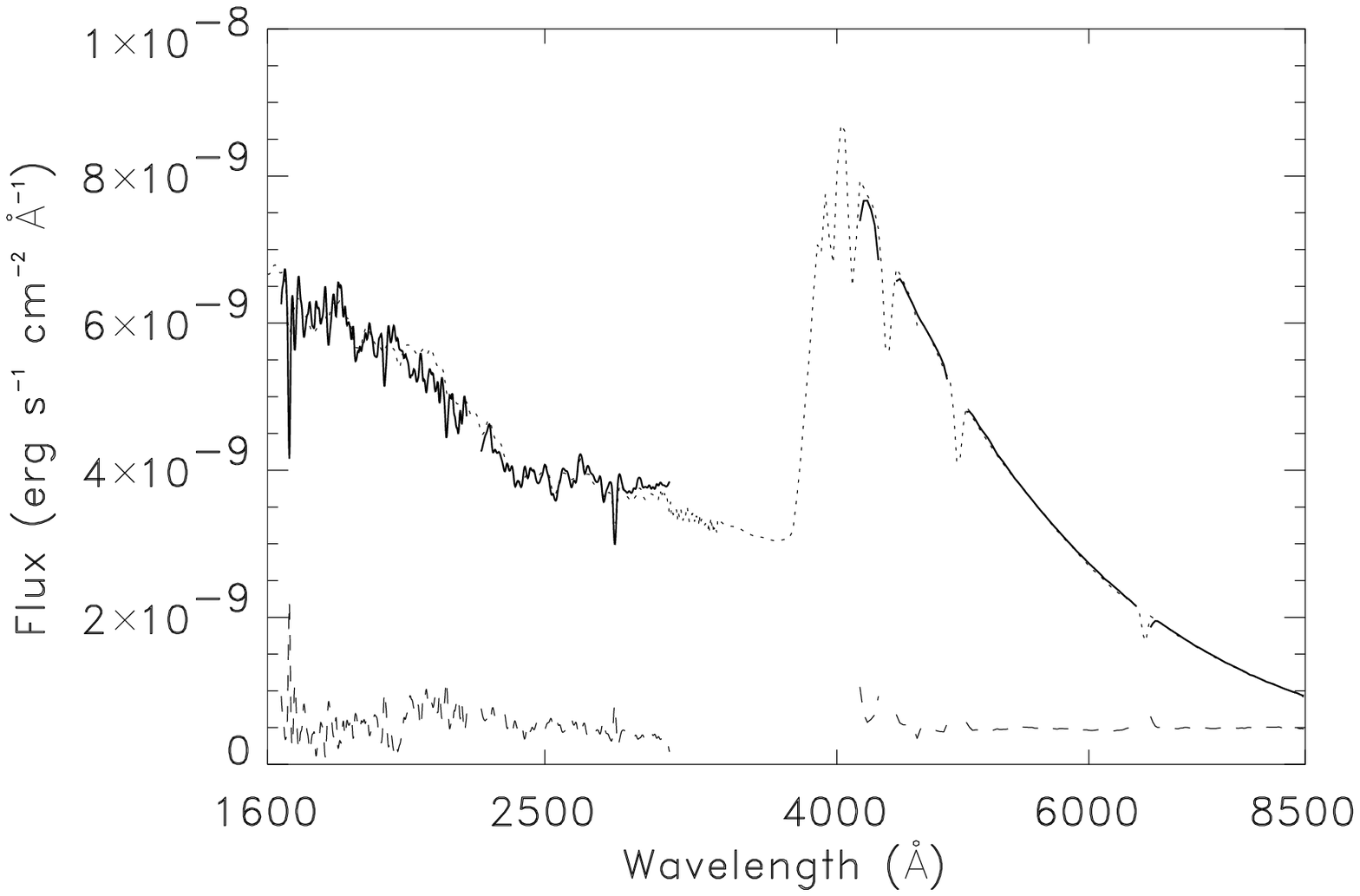} \par}
{\par \includegraphics[scale=0.7]{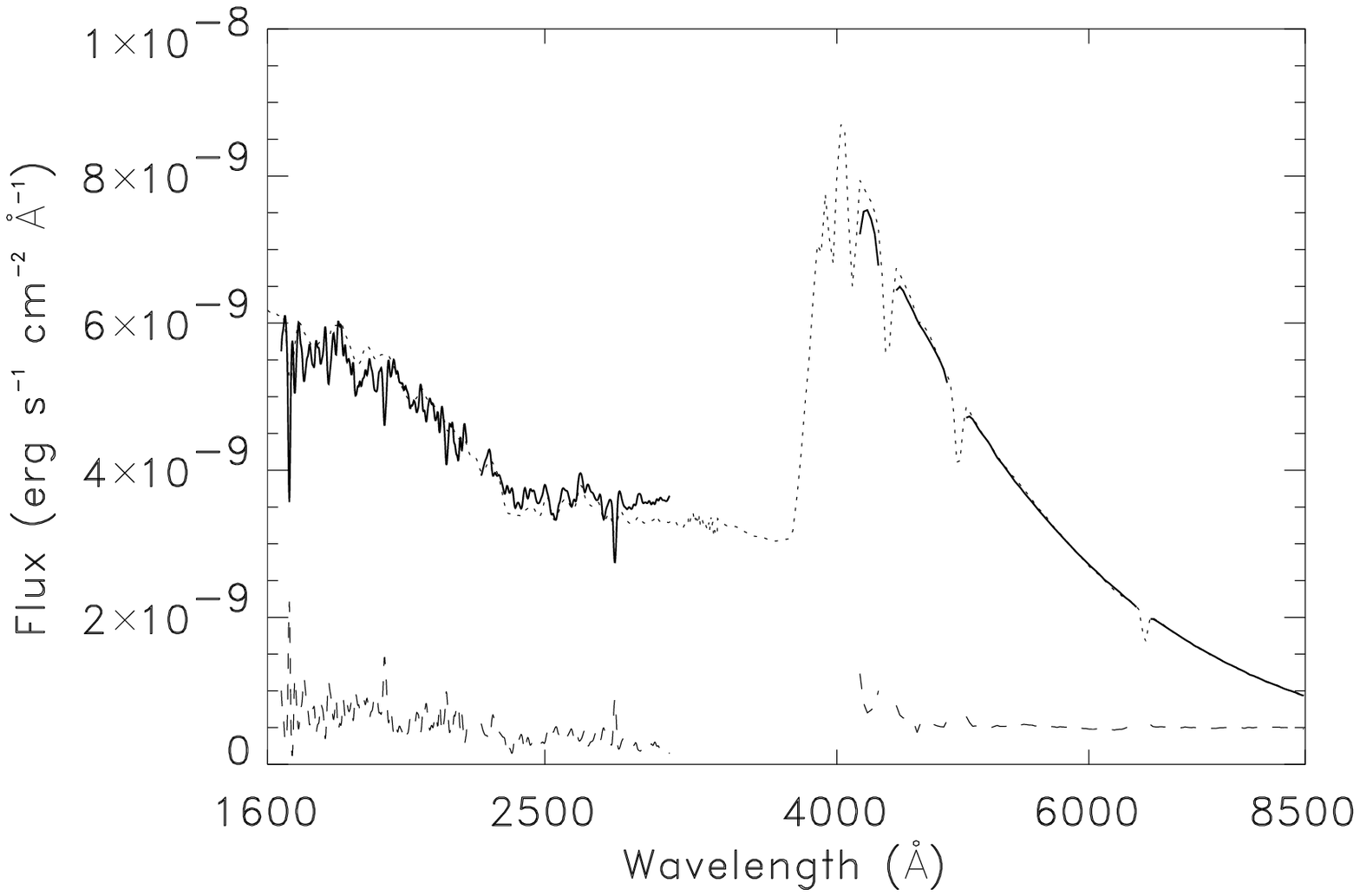} \par}
\end{center}
\caption{Best fit between the observed flux of Vega (dotted line) calibrated by 
Bohlin (top) and INES (bottom), and the computed flux (solid line). The residuals
(observed minus computed flux plus a constant) are also shown (dashed line). The stellar 
parameters associated with the adopted model atmosphere are shown in Table
\ref{fit_param}.}
 \label{fig:vis_spec}
\end{figure}

We minimized the sum of $\chi^2$ statistics between observed and computed
spectra in three wavelength regions weighted according to the approximate 
uncertainty in the observed fluxes inferred from Bohlin's analysis in the UV and 
Hayes' in the visible range, respectively (4\% between 1600 and 2150 \AA\ and 
between 2200 and 3000 \AA, and 2\% between 4150 and 8480 \AA, excluding Balmer 
lines). We used the Nelder-Mead simplex algorithm \citep{NM65} for the 
minimization. Using SYNSPEC, we computed a grid of fluxes based on interpolated 
Kurucz's model atmospheres \citep{K92} with T$_{\rm eff}$ running 
between 9350 and 9650 K. As the model
atmospheres are available in steps of 250 K, we linearly interpolated all
the relevant quantities as a function of the optical depth.
We matched selected spectral regions, excluding the segments 
where special difficulties exist. 
In particular, the spectrum in the vicinity of the Balmer jump
(3000--4150 \AA) was discarded 
because Hayes advises about significant errors in this region 
and the effect of fast rotation is important \citep{G94}. 
At this stage, a solar He abundance was used and the metal abundances 
are  from \citet{Q01}, including [Fe/H] = $-$0.57 dex, which
was adopted as the metallicity for the model atmosphere
[M/H] = $-$0.7 dex, using the same conversion factor as \citet{FM99}. 
 The best-fitting pairs (T$_{\rm eff}, \theta$) and the quality of the fit 
are displayed in Table \ref{fit_param}. The fit is clearly better in the UV using 
Bohlin's calibration than using the INES one, so the overall UV flux level is
better reproduced with synthetic spectra when the observed spectrum is in Bohlin's 
scale.
 Fig. \ref{fig:vis_spec} confronts the observed
flux (using Bohlin's calibration in the top panel
 and the INES calibration in the bottom panel) 
and the computed flux that best matches it. In that figure, the computed flux
is clearly lower than the observed one for $4150 < \lambda <$ 4500 \AA\ 
 when using the parameters of the best fit with INES
calibration, and it is clearly higher than the observed one for $2200 < \lambda 
<$ 3000 \AA. This can be explained when taking into account that INES UV flux 
is lower than Bohlin's one, but the same visible observed flux has been used.
Then, the computed flux matching the INES calibration is lower than the one
matching Bohlin's UV observed flux, and the same is observed 
in the visible next
to the UV region. Such effect is the result of diminishing
T$_{\rm eff}$ and increasing $\theta$ to match the INES scale.

A comparison between these parameters for Vega and those taken from 
the literature is made in Table \ref{param_liter}. The selected literature values
 do not include analyses of 
 UV spectra from IUE. 
 T$_{\rm eff}$ runs the gamut between 9430 and 9660 K 
 and $\theta$ between 3.24 and 
3.28 mas. These values are consistent with the parameters obtained with
 Bohlin's calibration (or with the M\&F one, given that both are on the 
same scale). However, 
the angular diameter derived from the analysis of the INES
spectrum is significantly higher than the literature values. In addition,
  there are only two T$_{\rm eff}$ determinations
that are consistent with 
the result we find if the INES calibration is adopted: 
the extreme value found
by  Moon \& Dworetsky (1985) and that by \citet{Ci01}. 
As the difference in the absolute flux between INES and the other 
calibrations amounts to more than 10\% for Vega, we have derived an average
UV spectrum as the mean of Bohlin's calibration and M\&F, not taking into account INES.
Using this flux, we have matched again the region between 2200 and 8500
\AA~(without the zone between 3000 and 4150 \AA), obtaining the following
result:

T$_{\rm eff}$ = 9620$^{+49}_{-63}$ K, $\theta$ = 3.272 $\pm$ 0.03 mas

\noindent where the
 error bars of our T$_{\rm eff}$ and $\theta$ estimations were derived 
by perturbing the best-fitting
parameters until the flux variation exceeded the error bars 
assigned to the observed spectrum.

The same analysis was carried out using new model atmospheres \citep{CK03}
which were calculated with 
overshooting turned off \citep[resulting in a steeper rise in temperature with depth in the
region where the optical continuum is formed; ][]{CGK97}, and using updated values for
the solar elemental abundances and revised opacity distribution functions (ODFNEW). 
The reason for this new analysis is to check the effect on colors of convective 
overshoot and the parameters thus obtained were: 

T$_{\rm eff}$ = 9575 K, $\theta$ = 3.271 mas,

\noindent where $\sigma_{2230}$ = 2.4 \% and $\sigma_{4085}$ = 0.7 \%. These 
values provide a slightly better fit, but are nonetheless consistent with 
thoses obtained using older model atmospheres
and also with most values from the literature.

\begin{figure}
   \epsscale{.80}
\plotone{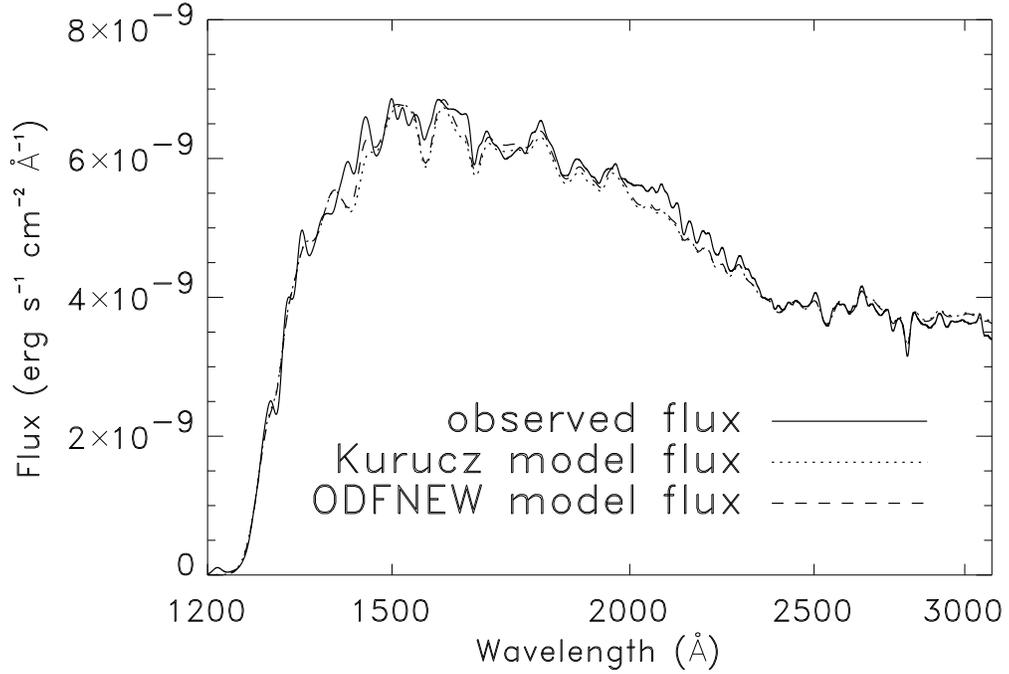}
\caption{Comparison between the observed spectrum of Vega 
(mean of Bohlin's and M\&F) and the computed spectrum using a Kurucz model atmosphere with
the following parameters: T$_{\rm eff}$ = 9620 K, $\log g$ = 3.98, [M/H] =
$-$0.7, [Si/H] = $-$0.90 and [C/H] = 0.03 and the computed spectrum using an ODF
model atmosphere with the following parameters: T$_{\rm eff}$ = 9575 K, $\log g$ = 3.98, [M/H] =
$-$0.7, [Si/H] = $-$0.90 and [C/H] = 0.03. The adopted angular diameter is 3.27
mas.}
 \label{fig:obs_comp} 
\end{figure}

Fig. \ref{fig:obs_comp} compares the observed and calculated fluxes,
including the interpolation of calculated fluxes when using ODFNEW model
atmospheres with these T$_{\rm eff}$ and $\theta$ values and C and Si
abundances found in section 3.4. There is a slight flux difference between both
synthetic spectra in the region between $\sim$ 1400 and $\sim$ 1900 \AA, where
the flux using ODFNEW models is higher, except $\sim$ 1560 \AA, where there is
no change. For instance, the fit is better $\sim$ 1450 \AA\ using ODFNEW models.

Adopting the  T$_{\rm eff}$ and [M/H] values from Alonso, 
Arribas \& Mart\'{\i}nez-Roger 
(1994), we searched only for the value of $\theta$ leading to the closest 
match to the observed spectrum in the near-UV and the visible ranges, 
obtaining again a value of 3.27 mas. Using this angular diameter, the fit to the
observed flux is very satisfying in the near-UV and visible spectral ranges, which
reinforces the validity of our method. We discuss the UV fluxes
below.

\begin{figure}
\begin{center}
{\par \includegraphics[scale=0.5]{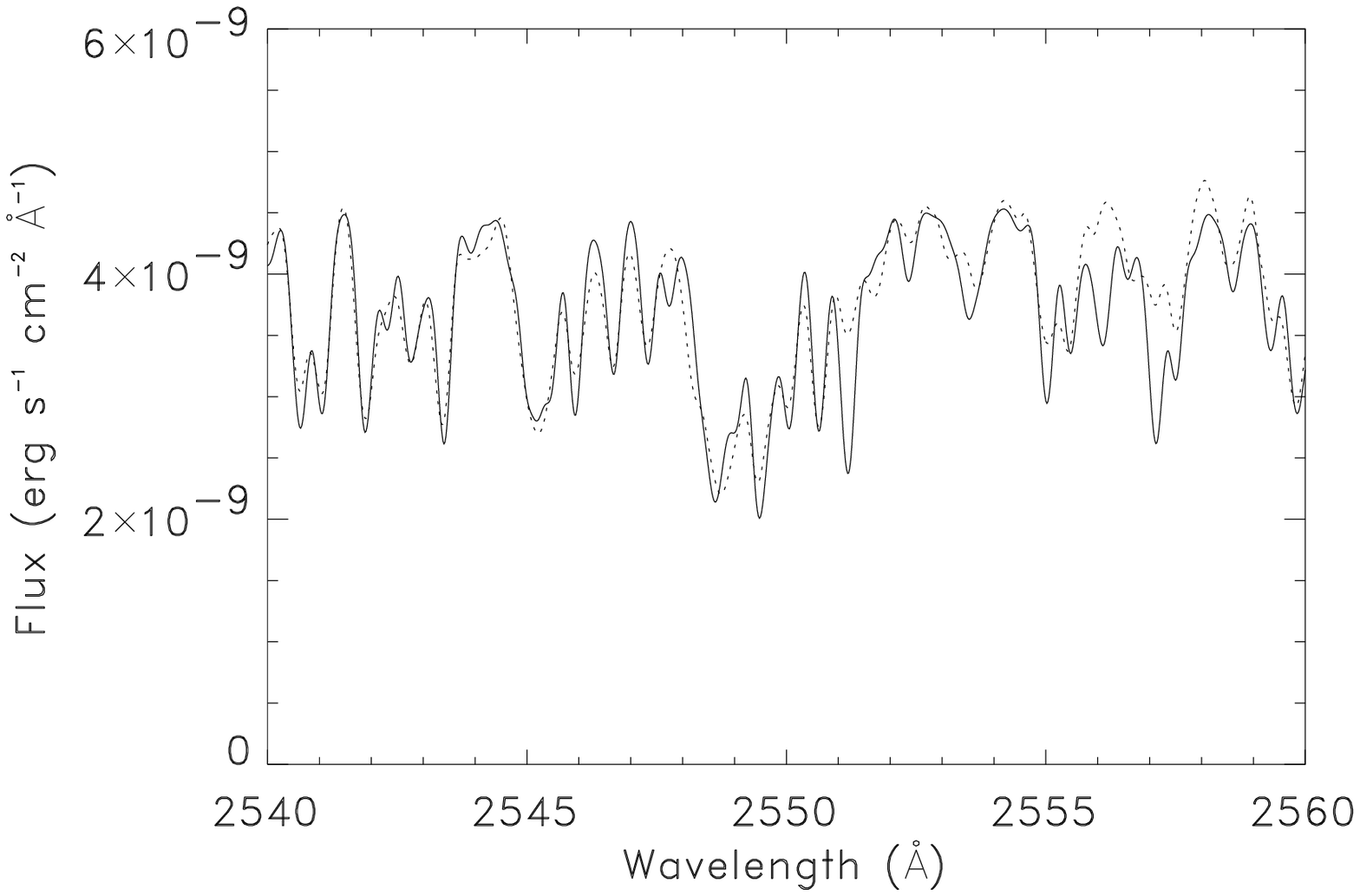} \par}
{\par \includegraphics[scale=0.5]{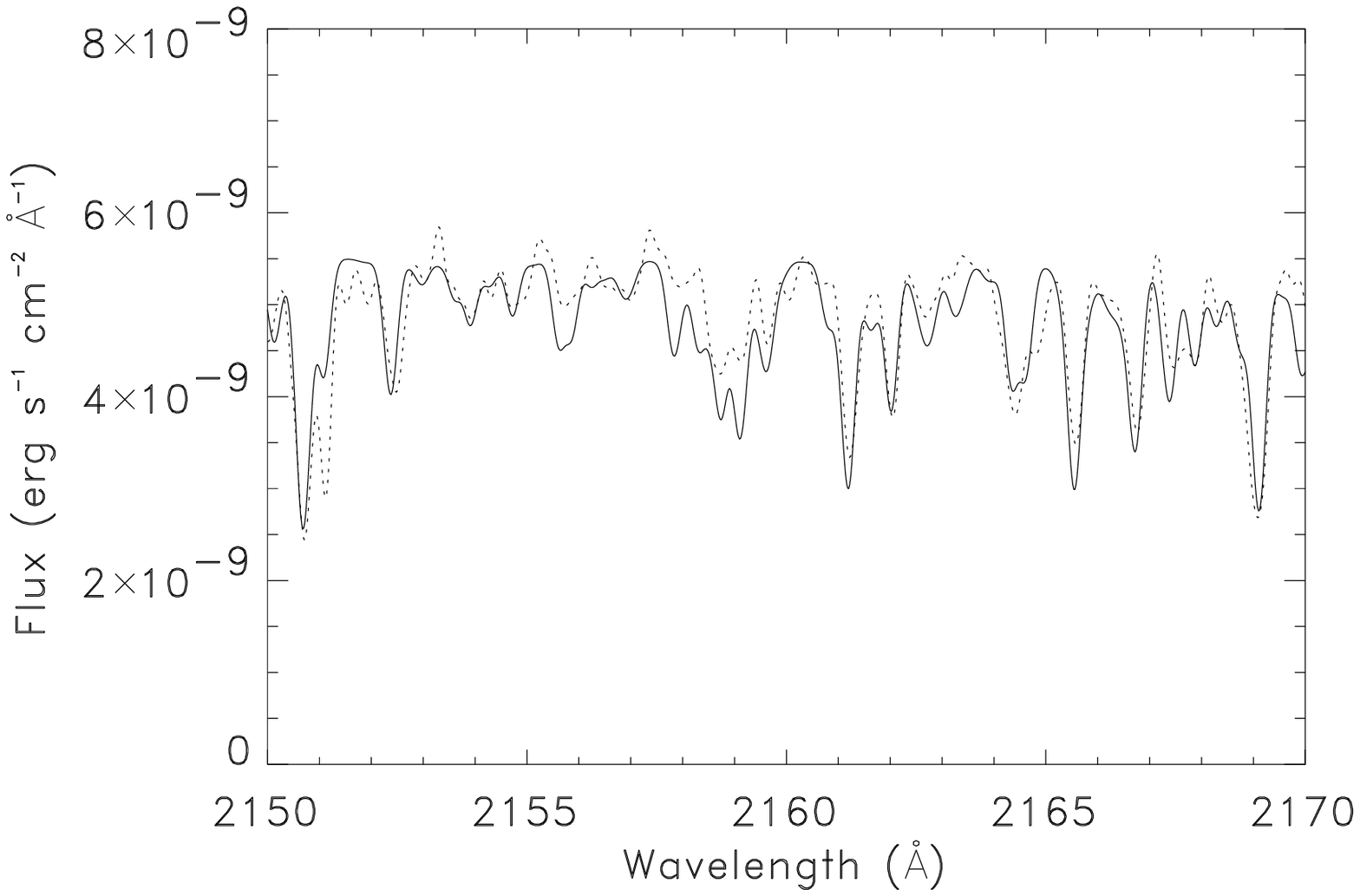} \par}
{\par \includegraphics[scale=0.5]{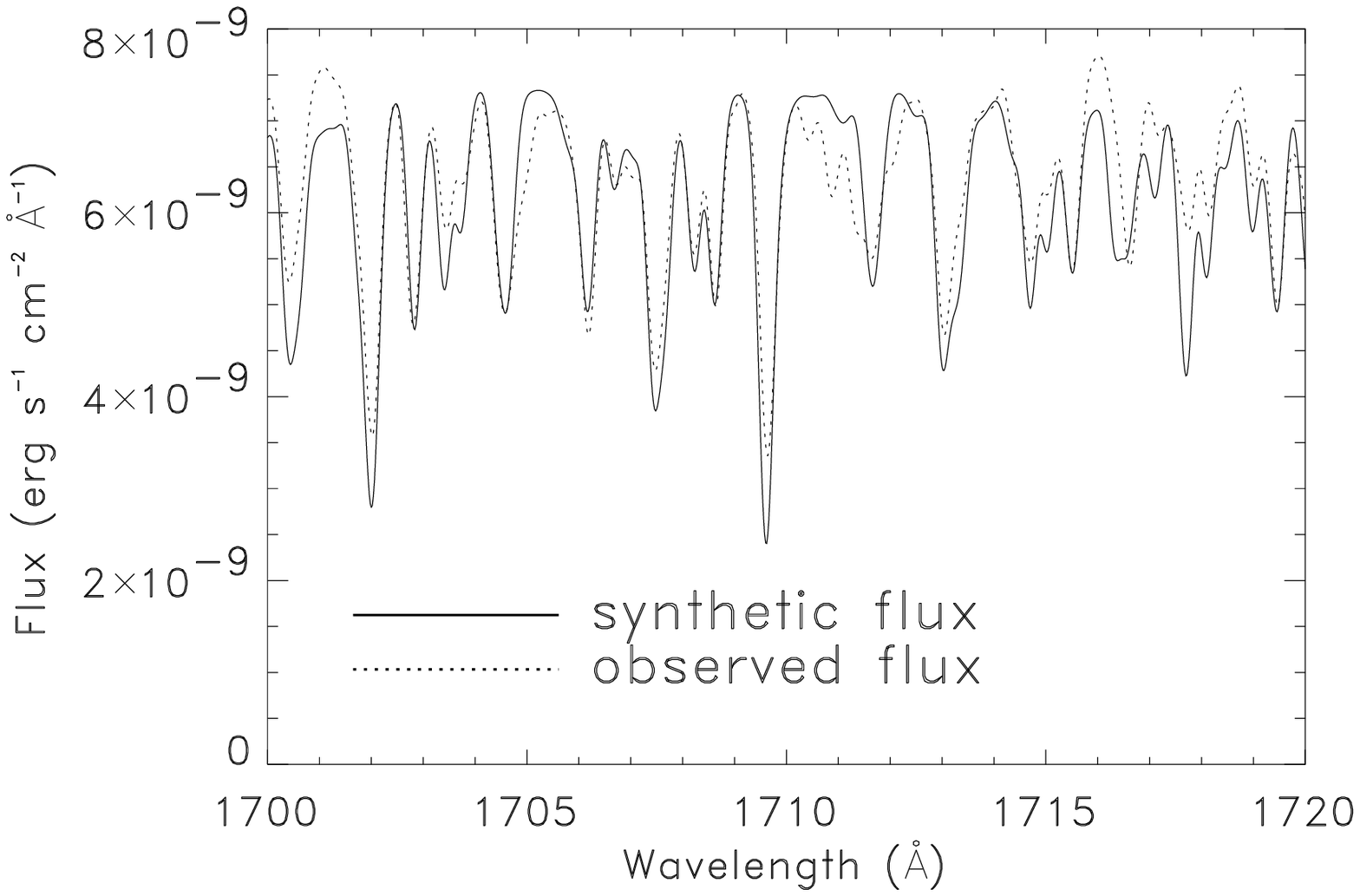} \par}
\end{center}
\caption{Comparison between the observed flux of Vega (dashed line, INES
multiplied by 1.1 to take the fluxes to the HST scale) and the computed 
flux (solid line) at high resolution. The stellar 
parameters associated with the adopted model atmosphere are T$_{\rm eff}$ = 9620 K
and log g = 3.98. The angular diameter used is 3.27 mas.}
 \label{fig:HR}
\end{figure}

A high-resolution spectrum using the interpolation of \citet{K92} model atmospheres 
with T$_{\rm eff}$ = 9620 K and log g = 3.98, and an angular diameter of 3.27 mas, 
has been computed and compared with the high-resolution IUE observed spectrum in 
three 100 \AA-wide spectral regions randomly selected
between 1550 and 3050 \AA, where the effects of rapid rotation and Si and C
continuous opacities on the flux are
very limited. The standard deviations between the computed and the observed spectra 
are: 8.5 \% in the region centered at 1668 \AA, 9.9 \% 
at 2174 \AA\ and 6.9 \% at 2550 \AA, which, although are higher than the values derived
from the low-resolution spectra, reproduce quite well the shape of the
high-resolution observations. Fig. \ref{fig:HR} compares the computed and the 
observed fluxes in three 20 \AA-wide windows, one in each region. 

\begin{figure}
\begin{center}
{\par \includegraphics[scale=0.7]{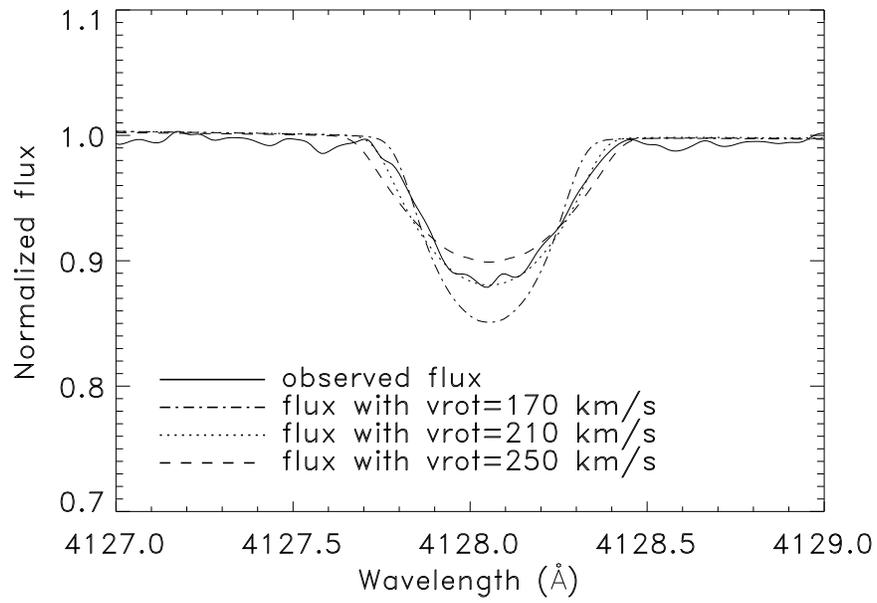} \par}
\end{center}
\caption{Comparison of the normalized high-resolution observed flux next to 
Si\,{\scriptsize II }$\lambda$4128.07 \AA\ and the normalized synthetic 
fluxes using T$_{\rm polar}$= 9595 K and i=6 degrees and different rotational
velocities: v$_{\rm rot}$(eq)= 170, 210 and 250 km/s.}
 \label{fig:rot}
\end{figure}

A preliminary test taking into account the rapid rotation on the emergent
flux of spectral lines was also performed. Some of the para\-me\-ters used (T$_{\rm polar}$ 
= 9595 K and i = 6 degrees) were taken from one of the best fits to the observed 
flux by \citet{G94}\footnote{However, the other parameters used are different. We 
have used the observed angular diameter (3.27 mas, taken from this paper) and the
distance (7.76 pc, obtained from the Hipparcos parallax) to derive the equatorial
radius (2.73 R$_{\rm solar}$), and also the polar radius by using the stellar
model. We have also used the mass of a typical A0V star (2.4 M$_{\rm solar}$) to
calculate the surface gravity.}, following the procedure used by 
\citet{PH99} to compute the flux. This procedure assumes uniform rotation and 
gravitational potential as a mass point. Using these stellar models with the line 
Si\,{\scriptsize II }$\lambda$4128.07 \AA, the synthetic line becomes deeper and
deeper and less and less broad when the rotational velocity decreases, obtaining a
good fit with v$_{\rm rot}$(eq) = 210 km/s, for which v sin i = 22.0 km/s, consistent with
the slow-rotating value. The comparison between the observed line and the computed
ones with v$_{\rm rot}$(eq) = 170, 210 and 250 km/s is shown in Fig. \ref{fig:rot}. This 
effect on the spectral lines will be considered in detail in future articles.

 \subsection{Estimation of Si and C abundances from the UV continuum}

\begin{figure}
\begin{center}
{\par \includegraphics[scale=0.7]{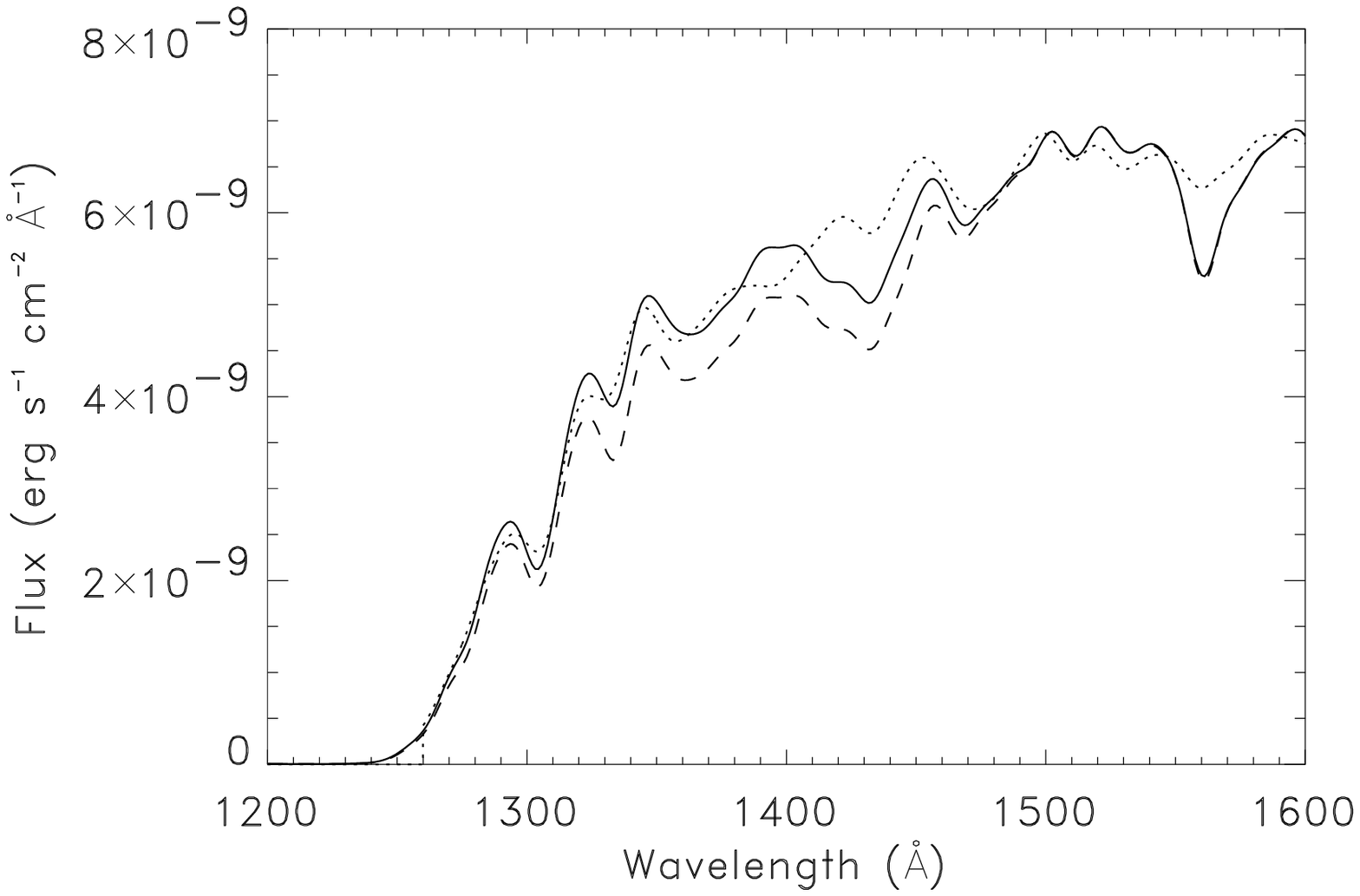} \par}
{\par \includegraphics[scale=0.7]{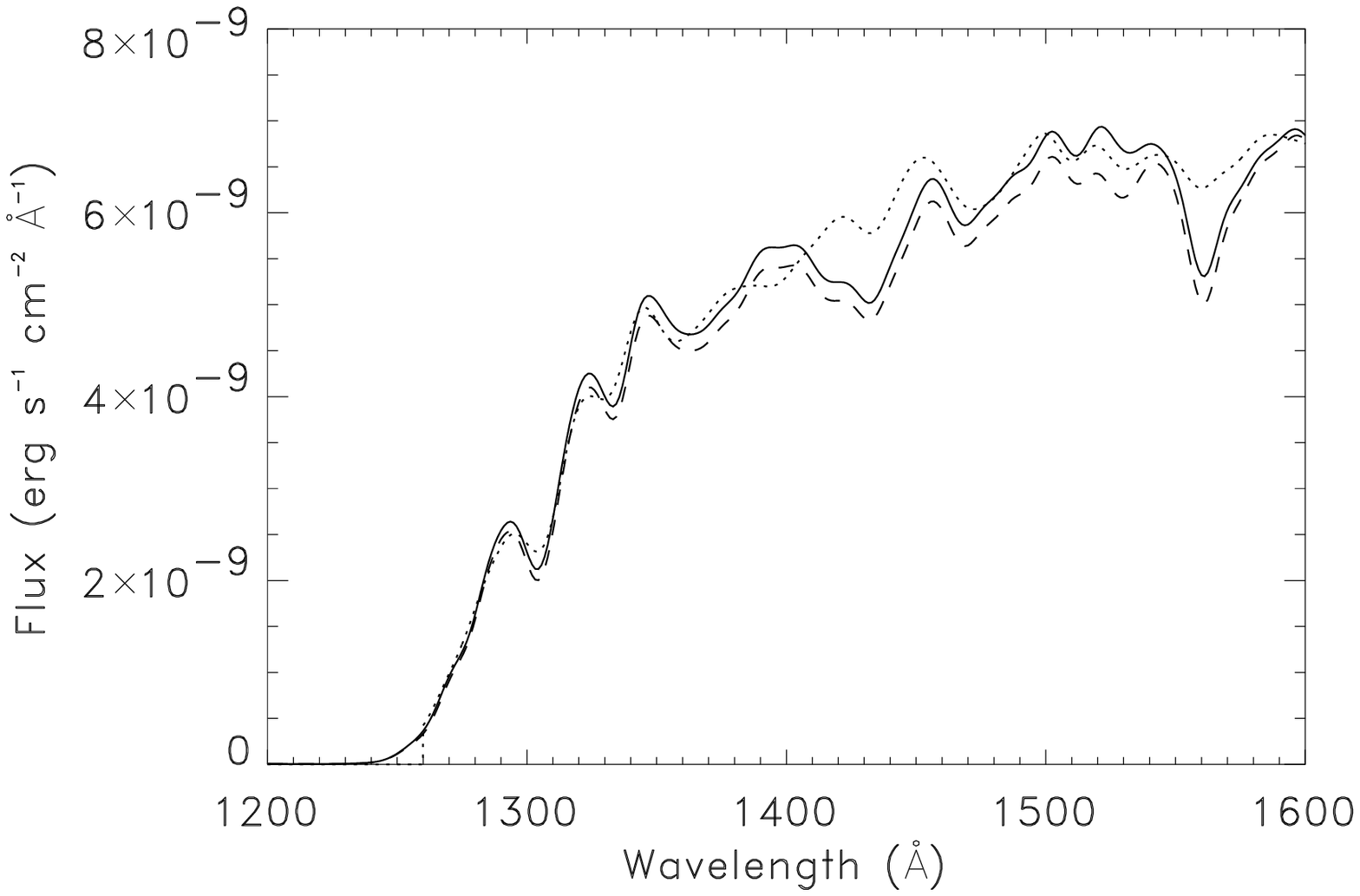} \par}
\end{center}
\caption{Comparison between the observed flux of Vega (dotted line) and the computed 
flux using [C/H] = +0.03 (solid line, top) and [C/H] = +0.23 dex (dashed line, top) and 
the computed flux using [Si/H] = $-$0.90 dex (solid line, bottom) and [Si/H] = $-$0.70 
dex (dashed line, bottom). The stellar 
parameters associated with the adopted model atmosphere are those obtained
in Section 3.3.}
 \label{fig:CSi}
\end{figure}

We turn to the main issue we are interested in, the analysis of
the UV spectrum.
In Section \ref{absorp}, we concluded that Si and C were important
contributors to the UV continuous opacity for a star like Vega.
With the adopted atmospheric parameters (gravity derived from the
{\it Hipparcos} parallax and both T$_{\rm eff}$ and $\theta$ from fitting
the optical and near-IR flux), we now attempt to constrain the 
abundances of these elements from the UV flux.
We calculated synthetic spectra from 1270 \AA~to 1380
\AA~and from 1460 \AA~to 1550 \AA.
The region between 1380 and 1460 \AA\ is not considered due to
exceptional problems that we discuss below.
We considered Si abundances \citep[from][] {GS98}
between [Si/H] = $-$1.1 and $-$0.5 
and C abundances between [C/H] = $-$0.1 and 0.2,
to find C and Si abundances.
In Fig. \ref{fig:CSi}, we can see the effect of a change  of 0.2 dex in the
abundances of C (top panel) and Si (lower panel).

The best fit to the observations is obtained for [Si/H] = $-$0.90 $^{+0.57}_{-2.10}$ 
and [C/H] = 0.03 $^{+0.34}_{-0.48}$, with $\sigma_{\rm rms}$ = 3\%, 
approximately the same as the uncertainty of the UV 
observed spectrum.
The error
bars take into account the uncertainty in He abundance (assumed $\pm$0.1 dex),
T$_{\rm eff}$, the observed flux, and Si abundance (C abundance) when computing 
the error in C abundance (Si abundance), where the first two error sources are the 
most relevant ones. The values obtained by
\citet{Q01} from the analysis of visible lines were 
[Si/H] = $-$0.59 $\pm$ 0.06 and 
[C/H] = $-$0.06 $\pm$ 0.13.
The reference solar abundances were adopted 
from the literature \citep{GS98} and not derived
by us following a similar procedure using the UV solar spectrum.

\begin{figure}
\begin{center}
{\par \includegraphics[scale=0.7]{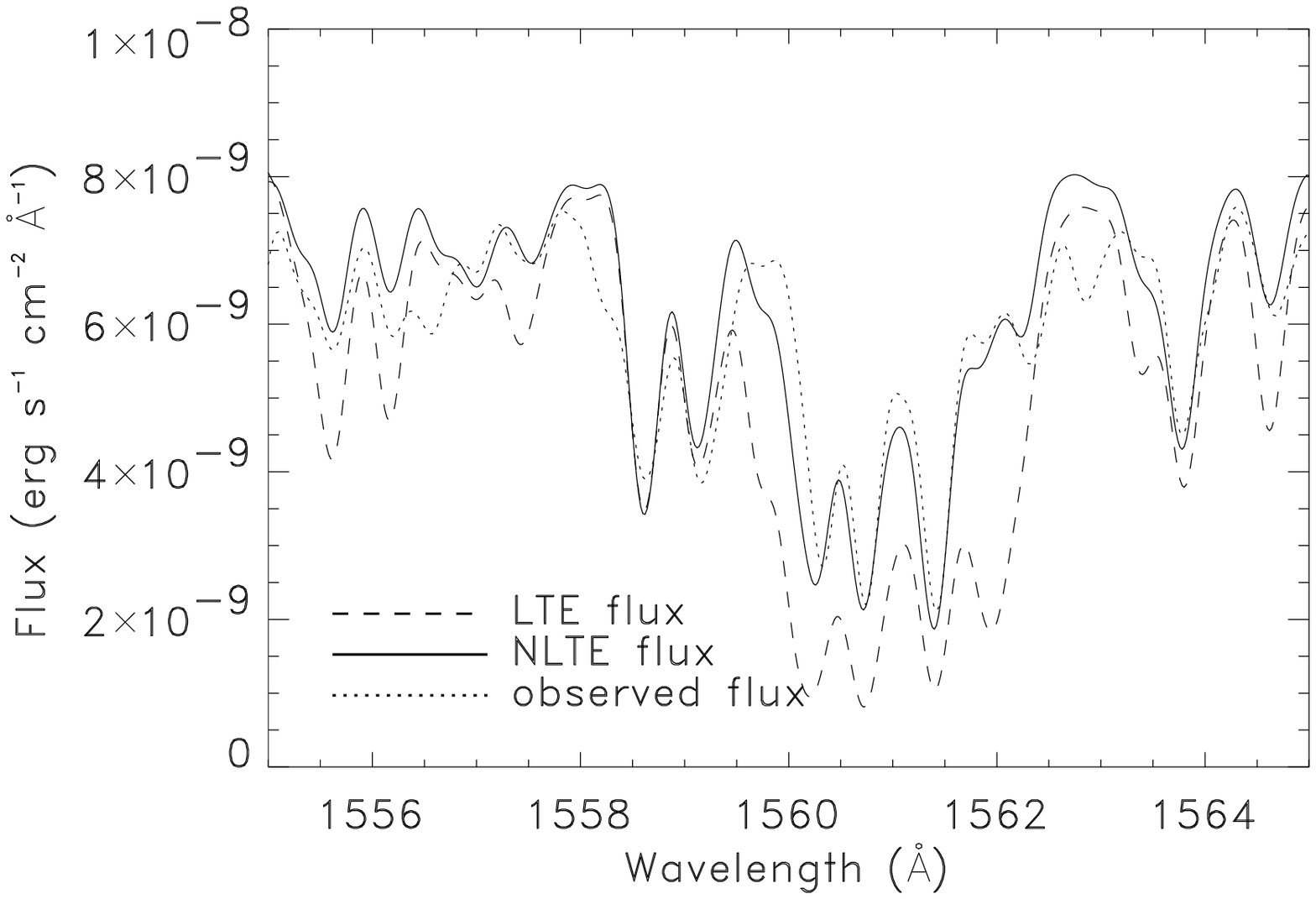} \par}
{\par \includegraphics[scale=0.7]{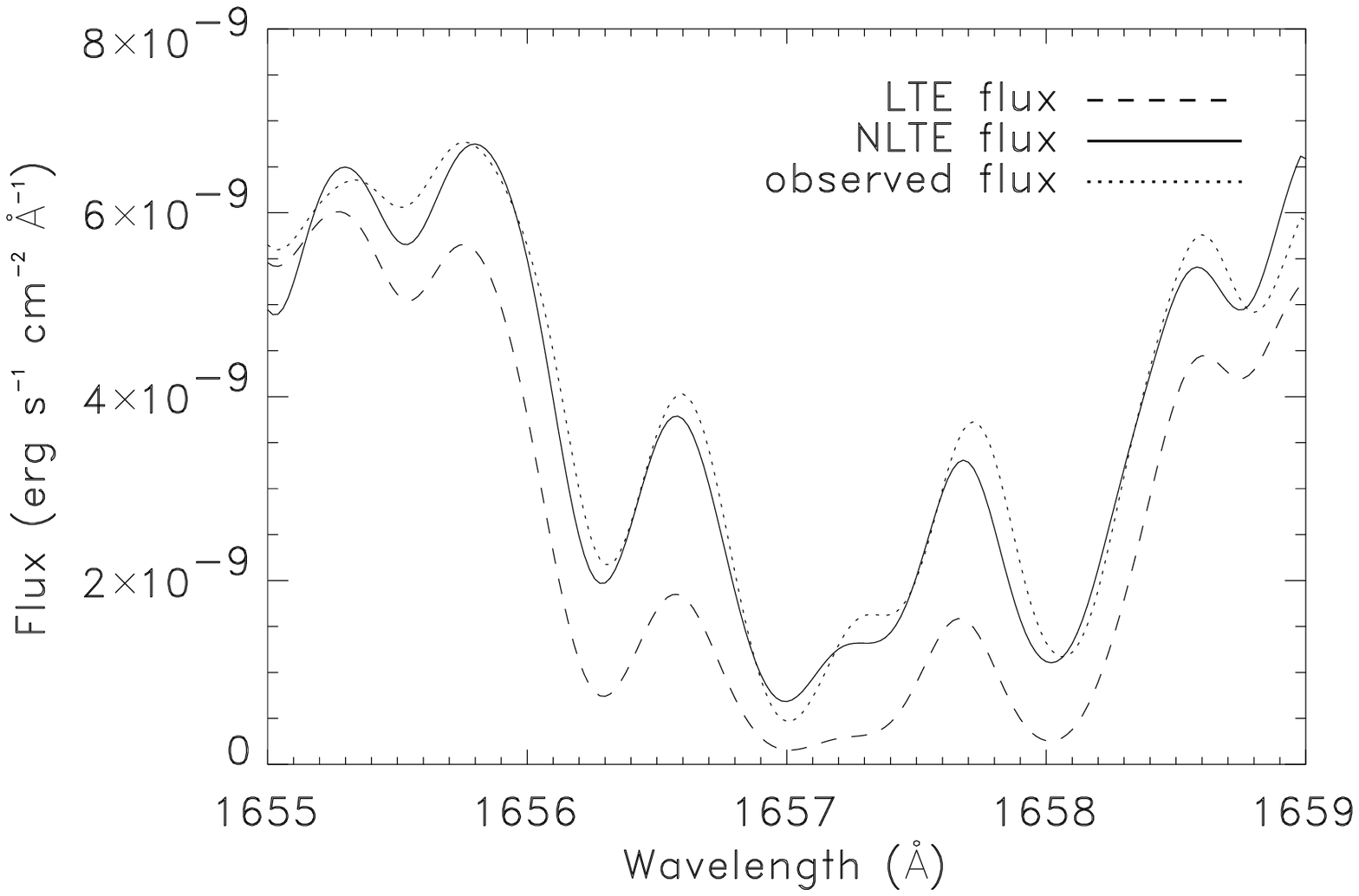} \par}
\end{center}
\caption{Comparisons between the observed spectrum of Vega 
(mean of Bohlin's and M\&F) and the computed spectrum next to two regions
(which have a poor fit in LTE in low resolution) using a Kurucz model atmosphere with
the following parameters: T$_{\rm eff}$ = 9620 K, $\log g$ = 3.98, [M/H] =
$-$0.7, [Si/H] = $-$0.90 and [C/H] = 0.03, in LTE and in NLTE. The regions
are $\sim$ 1560 \AA\ (top) and $\sim$ 1657 \AA\ (bottom). The adopted angular diameter is 3.27
mas.}
 \label{fig:NLTE} 
\end{figure}

The large error bars in the UV-based abundances are the result of
 the relatively
weak response of the UV continuum to changes in metal abundances, which 
is primarily a consequence of the dominant role of hydrogen opacity 
in such a warm  atmosphere. 
In addition, the existence of several spectral regions where the agreement
between observed and computed fluxes is, 
independently of the adopted abundances, poor, has quite a negative impact
on our potential to derive abundances from UV fluxes.
Some of the most remarkable dis\-cre\-pan\-cies are associated with
 deep Si\,{\scriptsize I} and C\,{\scriptsize I} lines (for instance, at
 $\sim$ 1560 \AA\ and at $\sim$ 1660 \AA), likely because departures
 from LTE, neglected in our calculations, are important. This effect could also
 affect the C and Si abundances derived from the UV continuum. However, some deep
 lines of other species (like Fe\,{\scriptsize II}) are well reproduced in
 LTE. A comparison between these C and Si lines in LTE and NLTE with the observed lines
 is shown in Fig. \ref{fig:NLTE}. There is a list of some of the deepest
 lines in these regions in Table \ref{NLTE_lines}.
 The regions between 1380 and 1460 \AA\ and between 
2000 and 2200 \AA\ show also a disappointing
mismatch between observed and calculated fluxes. 
 The reason for this failure is not clear (although the fit improves when using NLTE,
 but not enough), and a similar effect appears 
when using a revised flux of Vega based on STIS spectra (Bohlin, 
personal communication). These new fluxes agree much better with the computed flux in Balmer
lines, thus supporting the exclusion of these lines from our calculations. This
flux is lower than the one by Hayes, which would produce a lower T$_{\rm eff}$
and $\theta$ to match the computed flux.

  \subsection{Estimation of abundances by using spectral lines in the
 visible range}
 
To make a consistent comparison between UV abundances and those derived from
spectral lines in the visible region, we have selected a set of clean
 lines with a log gf uncertainty $\leq$ 25\% taken from 
\citet{Pr02}\footnote{We prefer to revisit the 
optical abundances for this star
because newer, high-quality data is now available, and because
that is the only way to guarantee the analysis is fully consistent with
our UV study. The reader should, however, keep in mind that our LTE analysis
is in some cases more limited than previously published studies such as
\citet{Pr02}, who studied several species in NLTE}, using their gf values, except
for Fe\,{\scriptsize II} lines, whose log gf values were taken from the adopted
values in \citet{AP02}.
We have excluded  from consideration He lines, because
they are too weak, and the Na\,{\scriptsize I} resonance 
doublet, because of the 
important departures from LTE in the core of these lines. 
We have measured the 
equivalent width (EW) of these lines by integrating the flux over the whole line 
using the {\it splot} procedure within
IRAF on the visible spectra taken at McDonald Observatory. These 
EW's were used to derive abundances with MOOG \citep{Sn02}. Classical line damping parameters
were used, that is, classical natural line broadening was assumed, as well as \"Unsold 
approximation for the Van der Waals broadening, while Stark broadening was 
neglected. A test about the influence of the Stark broadening using the VALD value on the derived 
abundances gives similar results as neglecting it.

Tables \ref{noFe_abun} and \ref{Fe_abun} show the observed EW 
of each spectral line (with its uncertainty) and the abundance  
(log N(X/H) + 12) derived using parameters 
in the model atmospheres from different references: \citet[][ {\it Qiu} in
Tables \ref{noFe_abun} and \ref{Fe_abun}] {Q01}, \citet[][ {\it Przybilla} in Tables 6 and 7] {Pr02} 
and ours (in which we have used a 
microturbulence of 2 km s$^{-1}$). 
Other reasonable values for the microturbulence (0-3 km s$^{-1}$) would
have a negligible effect on the mean iron abundance.
The mean difference of $\sim 0.1$ dex between iron abundances derived 
from
Fe\,{\sc i} and Fe\,{\sc ii} lines is perhaps due to the NLTE effects 
in the formation
of the former ones.

Using the abundances obtained from optical 
spectral lines, we found the mean 
abundance of each element, shown with its standard deviation in Table
\ref{abuns}. 
In the same table, we show the original abundances found in the literature 
using the
corresponding stellar parameters and, for C and Si,  
the values that we obtained from the analysis of the UV flux. Besides, solar abundances 
from \citet{GS98} are shown, written as GS98 in the table.
The uncertainties in the line-based 
abundances due to T$_{\rm eff}$ (9620$^{+49}_{-63}$K) and
to $\log g$ (assumed 3.98 $\pm$ 0.1) are in the range 0.01-0.07 dex 
and 0.01-0.06 dex, respectively.

We can see from Table \ref{abuns} that C, N, and O abundances for Vega are similar to
those of the Sun, while Mg, Al, Si, Ca, Cr, and Fe abundances are between
$-$0.3 and $-$0.7 dex compared to the solar ones. This is consistent with 
previous studies, and suggests that Vega is a mild $\lambda$ Boo star.

\section{Conclusions}

We have compiled spectroscopic observations from the literature to globally
examine the spectral energy distribution of Vega in the UV, optical and
near-IR. We also obtained new high-dispersion optical spectra and derived LTE
abundances from atomic lines to further constrain the modeling of the spectral
energy distribution.

Different versions of the mean UV spectrum based on IUE data are available,
and significant differences ($\sim 7$ \%) exist among them. We model Vega's
fluxes as a way to discriminate between the different flux scales. The
highest consistency is found adopting the HST flux scale. We identify the
most important contributors to the continuum opacity in the UV for Vega. The
bound-free opacities of neutral H and the H$^-$ ion play a dominant role 
above $\sim$1450 \AA. C\,{\scriptsize I} is the most important metal 
contributor below that wavelength, followed by Si\,{\scriptsize II}. A solar
He abundance has been assumed when calculating the observed spectrum,
consistent with some of the previous determinations \citep[e.g.][]{Pr02}, but
different to the lower value used by other authors \citep[e.g. in the
calculations by][]{G94}. This constitutes a non-negligible error source, since
a small change in that abundance produces a measurable variation in the
abundance of the H species, resulting in a significant change in the
continuum opacity or, equivalently, in the computed flux. 

The stellar effective temperature (T$_{\rm eff}$) and the angular diameter
($\theta$) are derived  by comparing computed spectra  with the observed
optical and near UV spectrum above 2200 \AA, where the only important 
contributors to the continuous opacity are H species, and metal abundances
are not very relevant. We obtain T$_{\rm eff}$ = 9620$^{+49}_{-63}$ K and
$\theta$ = 3.272 $\pm$ 0.03 mas. A good agreement between computed and observed
spectra is also obtained when using high resolution spectra in the UV, as
it is shown for three randomly selected regions. Once the atmospheric parameters were
derived, we made use of the UV flux to constrain the C and Si abundances. 

The abundance of Si obtained from the UV continuum  [Si/H] = $-0.90
^{+0.57}_{-2.10}$) is very different from the value we determine from the
spectral lines in the optical ([Si/H] = $-0.47 \pm 0.14$), but consistent
within the large uncertainty of the UV-based value. The abundance of C
obtained from the UV continuum ([C/H] = $+0.03^{+0.34}_{-0.48}$), however, is
in good agreement with the mean value determined from optical lines ([C/H] =
$-0.13 \pm 0.07$). Departures from LTE are known to affect both C\,{\scriptsize
I} and Si\,{\scriptsize II} lines \citep[Wedemeyer 2001;][]{SH90}, but are
small compared to the uncertainties of our UV-based abundances. We have 
shown that at least part of the spectral regions of the UV spectrum of Vega
that our LTE models cannot reproduce appear to be affected by departures from
LTE. In particular, two UV regions with deep C\,{\scriptsize I} and
Si\,{\scriptsize I} lines ($\sim$ 1560 and $\sim$ 1660 \AA) are better
reproduced when using NLTE models with C and Si abundances from the UV
continuum fit. A further study of high-dispersion spectra in the UV, departures from 
LTE in the modeling and the effect of rapid rotation is clearly the next step.

Together with a previous LTE analysis of the UV solar spectrum (Allende Prieto
et al. 2003), we arrive at the preliminary conclusion that our understanding of
stellar atmospheric opacities in the UV is fairly complete for spectral types
A to G. Departures from LTE may contribute somewhat for the solar case, but
are expected to be more important for a star like Vega (see, e.g., Hubeny
1991; Hauschildt et al. 1999). In the case of Vega, the distortion from
spherical symmetry induced by rapid rotation must be accounted for, in order
to disentangle such effect from departures from LTE and other approximations
adopted.

\clearpage

\clearpage

\clearpage

\clearpage

\clearpage

\clearpage

\clearpage

\clearpage

\clearpage

\clearpage

\begin{deluxetable}{c c c c c}
\tabletypesize{\scriptsize}
\tablecaption{Spectra from different IUE web pages averaged to calculate the 
calibrated spectrum of Vega with different dispersion and wavelength ranges. M\&F fluxes
are obtained from the MAST-IUE web page and INES fluxes from the INES web 
page.\label{IUE_cam}}
\tablewidth{0pt}
\tablehead{
  \colhead{calibration}&\colhead{dispersion}&\colhead{camera name}&\colhead{wavelength range}&\colhead{spectrum numbers}
}
\startdata
  M\&F and INES&low&SWP&1150-2000 \AA&27024, 29864, 30548, 30549, 32906, 32907\\
  M\&F and INES&low&LWP&1850-3300 \AA&07010, 07038, 07888, 07889, 07902, 07903, 07904, 10347, 10348\\ 
  INES&high&SWP&1150-2000 \AA&01188, 24504, 32870, 32887, 33633, 38411, 42521, 45285 \\
  INES&high&LWR&1850-3300 \AA&07585, 08630\\
  INES&high&LWP&1850-3300 \AA&07490, 12617, 17648, 21295, 23642\\ 
\enddata
\end{deluxetable}
\clearpage

\begin{deluxetable}{c c c c c c}
\tabletypesize{\scriptsize}
\tablecaption{Parameters and errors of the best fit between synthetic and observed
  spectra of Vega using different calibrations.\label{fit_param}}
\tablewidth{0pt}
\tablehead{
  \colhead{calibration}&\colhead{T$_{\rm eff}$ (K)}&\colhead{$\theta$
  (mas)}&\colhead{$\sigma_{1622}$}&\colhead{$\sigma_{2230}$}&\colhead{$\sigma_{4085}$}
}
\startdata
  Bohlin's&9631&3.272&4.9 \%&2.9 \%&1.3 \%\\
  INES&9458&3.324&5.9 \%&5.7 \%&1.4 \%\\ 
\enddata
\end{deluxetable}
\clearpage

\begin{deluxetable}{c c c c}
\tabletypesize{\scriptsize}
\tablecaption{Parameters for Vega according to the 
literature.\label{param_liter}}
\tablewidth{0pt}
\tablehead{
  \colhead{Reference}&\colhead{T$_{\rm
  eff}$ (K)}&\colhead{$\theta$ (mas)}&\colhead{[M/H] (dex)}
}
\startdata
Code et al. (1976)\tablenotemark{1}&9660 $\pm$ 140&3.24 $\pm$ 0.07&-\\
Castelli \& Kurucz (1994)\tablenotemark{2}&9550&-&$-$0.5\\
Ciardi et al. (2001)\tablenotemark{3}&9553 $\pm$ 111&3.28 $\pm$ 0.01&-\\
Fitzpatrick \& Massa (1999)\tablenotemark{4}&9600 $\pm$ 10&3.26 $\pm$ 0.01&$-$0.73 $\pm$
0.06\\
Moon \& Dworetsky (1985)\tablenotemark{5}&9430 $\pm$ 200&-&-\\
Alonso, Arribas \& Mart\'{\i}nez-Roger (1994)\tablenotemark{6}&9610&3.24 $\pm$ 0.07&$-$0.25 \\
Mozurkewich et al. (2003)\tablenotemark{7}&9657 $\pm$ 119&3.225 $\pm$ 0.032&-\\
This paper\tablenotemark{8}&9620$^{+61}_{-56}$&3.27 $\pm$ 0.03&$-$0.7\\ 
\enddata

\tablenotetext{1}{Obtained from angular size measurements and its absolute
flux distribution}

\tablenotetext{2}{Parameters of the Kurucz model producing the synthetic spectrum that
best matches the observed one derived from the model to the observed one (in the
visible and the UV, with Bohlin's calibration, and near-IR). A solar 
abundance of He has been adopted, with no reddening and with a microturbulence
of 2 km s$^{-1}$}

\tablenotetext{3}{Angular diameter obtained from interferometry at 2.2 $\mu$m}

\tablenotetext{4}{The uncertainties given are 1$\sigma$ internal errors 
for the fittings of the synthetic spectrum derived from Kurucz models to the
observed one (in the visible, the near-IR, and the near-UV, with the M\&F
calibration) }

\tablenotetext{5}{Based on uvby$\beta$ photometry, with error estimation
from Qiu et al. (2001)}

\tablenotetext{6}{Using the IFM to calibrate the
absolute flux of stars with direct measurements of angular size}

\tablenotetext{7}{Angular diameter obtained from optical interferometry}

\tablenotetext{8}{Using Kurucz (1992) model atmospheres}

\end{deluxetable}

\clearpage

\begin{deluxetable}{c c}
\tabletypesize{\scriptsize}
\tablecaption{Some of the deepest lines in the regions shown in Fig. \ref{fig:NLTE}
($\sim$ 1560 and $\sim$ 1657 \AA), where a study of the NLTE effect has been
carried out.\label{NLTE_lines}}
\tablewidth{0pt}
\tablehead{
  \colhead{$\lambda$ (\AA)} &\colhead{ion}\\
}
\startdata
 1555.660&Si I\\
 1556.161&Si I\\
 1558.541&Fe II\\
 1558.692&Fe II\\
 1559.085&Fe II\\
 1559.706&Si I\\
 1560.309&C I\\
 1560.682&C I\\
 1560.709&C I\\
 1561.340&C I\\
 1561.438&C I\\
 1562.002&Si I\\
 1563.790&Fe II\\
 1564.612&Si I\\
\tableline 
 1656.267&C I\\
 1656.929&C I\\
 1657.008&C I\\
 1657.379&C I\\
 1657.907&C I\\
 1658.121&C I\\
\enddata
\end{deluxetable}

\clearpage

\begin{deluxetable}{c c c c c c}
\tabletypesize{\scriptsize}
\tablecaption{Chemical abundances obtained from spectral lines in the visible
  region.\label{noFe_abun}}
\tablewidth{0pt}
\tablehead{
  \colhead{ion}&\colhead{$\lambda$ (\AA)}&\colhead{EW
  (m\AA)}&\colhead{log(N/X) with} 
  &\colhead{log(N/X) with} &
  \colhead{log(N/X)} \\
  & & & \colhead{parameters from Qiu}&\colhead{parameters from Przybilla}&
  \colhead{with our parameters}
}
\startdata
 C I  & 5052.17&       33$\pm$4&        8.37& 8.44&    8.47\\
      & 5380.34&       20$\pm$3&        8.30& 8.37&    8.39\\
      & 6828.12&        7$\pm$2&        8.25& 8.32&    8.34\\
      & 7111.47&       17$\pm$5&        8.35& 8.42&    8.44\\
      & 7113.18&       23$\pm$2&        8.19& 8.26&    8.28\\
      & 7115.17&       23$\pm$2&        8.34& 8.41&    8.44\\
      & 7116.99&       21$\pm$2&        8.28& 8.35&    8.37\\  
\tableline
 N I&7423.64 &20.5$\pm$3   &8.11 &8.12 &8.12\\
    &7442.30 &34$\pm$3 &8.08 &8.08& 8.09\\
    &7468.31 &42$\pm$4   &8.05 &8.04& 8.05\\
\tableline
 O I&5329.68\tablenotemark{1} &23$\pm$4& 8.77 &8.77& 8.77\\
    &6158.19\tablenotemark{2} &57$\pm$3& 8.81 &8.79& 8.80\\
\tableline
 Mg I&4167.27\tablenotemark{3}& 20$\pm$3   & 7.10& 7.19&7.22\\
     &4702.99    & 30$\pm$3   & 6.96& 7.04&7.07\\
     &5183.60\tablenotemark{4}& 122.5$\pm$2& 7.40& 7.31&7.35\\
     &5528.41    &  28$\pm$2& 6.92& 7.00&7.03\\
\tableline
 Mg II&4427.99& 5$\pm$2&  6.88& 6.87& 6.87\\
      &4433.99&12$\pm$2&6.97& 6.95& 6.95\\
\tableline
 Al II&4663.05& 4.5$\pm$2& 5.81& 5.77& 5.77\\
\tableline
 Si II &4128.07    &  54$\pm$4	&7.02& 6.97& 6.96\\
      &4130.89    &  68$\pm$4	&6.85& 7.04& 7.04\\
      &6347.11\tablenotemark{5}& 114$\pm$4	&7.63& 7.49& 7.42\\ 
      &6371.37    & 80$\pm$3   &7.30& 7.22& 7.23\\
\tableline
 Ca I  &4226.73      &58$\pm$3& 5.62& 5.71& 5.76\\
\tableline
 Cr I  &5206.02&      11$\pm$2&      5.24& 5.35& 5.39\\
       &5208.42&      15$\pm$2&      5.26& 5.37& 5.41\\

\enddata
\tablenotetext{1}{This is a set of several weak OI lines according to the line list
taken from SYNPLOT. However, \citet{Pr02} seems to take an effective gf with the
central wavelength of the table, which is what we have used}

\tablenotetext{2}{Comparing the computed triplet with the observed "line"}

\tablenotetext{3}{Using the log gf of \citet{Q01}: $-$0.79; using the log gf of
Przybilla's PhD Thesis (2002), $-$1.71, it is obtained a value 0.92 dex
higher}

\tablenotetext{4}{Important NLTE effect}

\tablenotetext{5}{The abundances using damping parameters from VALD are very different, so
we have not used this line to derive the abundance of Si}

\end{deluxetable}
\clearpage

\begin{deluxetable}{c c c c c c}
\tabletypesize{\scriptsize}
\tablecaption{Chemical abundances for Fe obtained from spectral lines in the visible
  region.\label{Fe_abun}}
\tablewidth{0pt}
\tablehead{
  \colhead{ion}&\colhead{$\lambda$ (\AA)}&\colhead{EW (m\AA)}&\colhead{log(N/X) with}
   &\colhead{log(N/X) with} &\colhead{log(N/X) with} \\
  & & &\colhead{parameters from Qiu}&\colhead{parameters from Przybilla}&
  \colhead{our parameters}
}
\startdata
 Fe I &4187.04     & 11$\pm$2     & 6.74             & 6.84 &6.87\\
      &4187.79     & 16$\pm$3       & 6.92             & 7.02 &7.05\\
      &4202.03     & 31$\pm$3     & 6.89             & 6.97 &7.01\\
     & 4404.75     & 53.5$\pm$3       & 6.81             & 6.84 &6.88\\
      &4459.12     &  7$\pm$2       & 7.06             & 7.16 &7.19\\
      &4494.56     &  8$\pm$2       & 6.99             & 7.09 &7.13\\
     & 4736.77     &  5.5$\pm$2       & 7.04             & 7.14 &7.17\\
     & 5324.18     & 15$\pm$2       & 7.02             & 7.11 &7.14\\
     & 5367.47     &  8$\pm$2       & 6.89             & 6.98 &7.01\\
     & 5369.96     & 12$\pm$2     & 7.02             & 7.10 &7.13\\
     & 5383.37     & 14$\pm$2     & 6.92             & 7.01 &7.04\\
     & 5410.91     &  9$\pm$2       & 7.00             & 7.09 &7.12\\
     & 5569.62     &  5$\pm$1     & 6.94             & 7.03 &7.07\\
     & 5572.84     &  7$\pm$2       & 6.85             & 6.95 &6.98\\
     & 5586.76     & 11$\pm$2     & 6.95             & 7.05 &7.08\\
     & 5615.64     & 13$\pm$2       & 6.94             & 7.03 &7.06\\
\tableline
 Fe II &       4173.47&  61$\pm$3  &	 7.01&     6.94&     6.95\\
       &       4178.87&  59$\pm$3  &	 6.93&     6.86&     6.87\\
       &       4233.17& 103$\pm$4 &	7.45&	  7.15&     7.17\\
       &       4303.17&  42.5$\pm$5   &     6.82&     6.80&	6.81\\
       &       4351.76&  73$\pm$5   &	  6.88&     6.74&     6.76\\
       &       4385.39&  48$\pm$4 &	7.06&	  7.03&     7.04\\
       &       4491.40&  39$\pm$3   &	  6.89&     6.88&     6.89\\
       &       4508.29&  58$\pm$3   &	  7.13&     7.07&     7.09\\
       &       4515.33&  50$\pm$4 &	6.78&	  6.75&     6.76\\
       &       4522.62&  71$\pm$5 &	6.88&	  6.76&     6.77\\
       &       4555.88&  61.5$\pm$4   &     6.92&     6.85&	6.86\\
       &       4576.33&  27$\pm$4 &	6.87&	  6.88&     6.89\\
       &       4582.83&  20.5$\pm$3   &     6.87&     6.88&	6.89\\
       &       4583.83&  93$\pm$5   &	  7.18&     6.94&     6.95\\
       &       4620.51&  14.5$\pm$3   &     6.80&     6.82&	6.83\\
       &       4629.34&  56$\pm$3   &	  6.82&     6.76&     6.77\\
       &       4656.97&  11$\pm$3 &	7.09&	  7.12&     7.13\\
       &       4923.92&  118.5$\pm$5   &   7.37&     7.06&     7.08\\	    
       &       5018.45&  128.5$\pm$3  & 7.55&	   7.24&    7.28\\
       &       5169.00&  134$\pm$3  &	7.31&	   7.01&    7.05\\
       &       5197.57&  54$\pm$4   &	  6.81&     6.77&     6.78\\
       &       5234.66&  60$\pm$5   &	  7.11&     7.04&     7.05\\
       &       5276.00&  64$\pm$5   &	  6.86&     6.78&     6.79\\
       &       7711.73&  12$\pm$3 &	6.70&	  6.72&     6.73\\ 
\enddata
\end{deluxetable}
	\clearpage

\begin{deluxetable}{c c c c c}
\tabletypesize{\scriptsize}
\tablecaption{Chemical abundances obtained from spectral lines in the visible
  compared with those from the literature, from the UV region, and for the 
  Sun.\label{abuns}}
\tablewidth{0pt}
\tablehead{
  \colhead{ion}& \colhead{log N(X) with} &\colhead{log N(X)
  with}&\colhead{log N(X) with} & 
  \colhead{log N(X) for the Sun}\\
  &\colhead{parameters from Qiu}&\colhead{parameters from Przybilla}&
  \colhead{our parameters}&\colhead{from GS98}
}
\startdata
  C I& & & &\\
  mean&8.30$\pm$0.07&8.37$\pm$0.06&8.39$\pm$0.07&\\
  literature & 8.46$\pm$0.13 &8.23$\pm$0.11& &8.52\\
 UV continuum & & &8.55 $^{+0.34}_{-0.48}$&\\
\tableline
 N I& & & &\\
 mean	   & 8.08$\pm$0.03& 8.08$\pm$0.04&8.09$\pm$0.04&\\ 
 literature     & 8.00$\pm$0.02&7.69$\pm$0.06 (NLTE)& &7.92\\ 
\tableline
 O I& & & &\\
    mean           &8.79$\pm$0.03 &  8.78$\pm$0.02 &8.78$\pm$0.02 & \\   
    literature	   &9.01$\pm$0.14&  8.57$\pm$0.05 (NLTE)& &8.83\\
\tableline
 Mg I& & & &\\
    mean	  &7.10$\pm$0.22 &7.14$\pm$0.14	  &7.17$\pm$0.15&\\
    literature     &6.81 (1 line)&7.02$\pm$0.06 (NLTE)& &7.58\\
\tableline
 Mg II& & & &\\
    mean	   &6.92$\pm$0.06& 6.91$\pm$0.06&  6.91$\pm$0.06&\\
    literature     &6.69$\pm$0.05& 7.02$\pm$0.03& &7.58\\ 
\tableline
 Al II& & & &\\
   mean &5.81 (1 line)&5.77 (1 line)&5.77 (1 line)&\\
   literature     &---& 5.84 (1 line)& &6.47\\  
\tableline
 Si II & & & &\\
     mean	    &7.06$\pm$0.23&7.08$\pm$0.13& 7.08$\pm$0.14&\\
     literature     &6.96$\pm$0.06& 6.94$\pm$0.05& &7.55\\ 
     UV continuum   & & &6.65 $^{+0.57}_{-2.10}$\\
\tableline
 Ca I  & & & &\\
   mean & 5.62 (1 line)& 5.71 (1 line)& 5.76 (1 line)&\\
   literature     &5.41$\pm$0.09& 5.67$\pm$0.11& &6.36\\
\tableline
 Cr I  & & & &\\
    mean	   & 5.25$\pm$0.01& 5.36$\pm$0.01&5.40$\pm$0.02&\\
    literature     & 5.19$\pm$0.17 (CrII)  &5.12$\pm$0.04& &5.67\\ 
\tableline
 Fe I & & & &\\
     mean	&6.94$\pm$0.09& 7.03$\pm$0.10& 7.06$\pm$0.09&\\
     literature	&6.94$\pm$0.12& 6.96$\pm$0.08& &7.50\\ 
\tableline
 Fe II & & & &\\
     mean	&7.00$\pm$0.23& 6.91$\pm$0.15& 6.92$\pm$0.15&\\
     literature &6.93$\pm$0.13& 6.97$\pm$0.12 (NLTE)& &7.50\\  
\enddata
\end{deluxetable}

\end{document}